\documentclass[aps,showpacs,twocolumn,longbibliography,nofootinbib,secnumarabic,eqsecnum]{revtex4-2}

\usepackage{graphicx}
\usepackage{dcolumn}
\usepackage{bm}
\usepackage{braket}
\usepackage{leftidx}
\usepackage{cancel}
\usepackage{amsmath}
\usepackage{epstopdf}
\usepackage{color}
\usepackage[mathcal]{eucal}
\usepackage{float}

\usepackage{yfonts}
\usepackage{color}

\usepackage{scalerel}
\usepackage{mathtools}

\begin{document}
\date{\today}
\title{Anatomy of entanglement and nonclassicality criteria}

\author{Mehmet Emre Tasgin}

\affiliation{Institute of Nuclear Sciences, Hacettepe University, 06800, Ankara, Turkey}
\affiliation{metasgin@hacettepe.edu.tr and metasgin@gmail.com}

\begin{abstract}
We examine the internal structure of two-mode entanglement criteria for quadrature- and number-phase-squeezed states. For criteria obtained from the partial transpose of the Schr\"odinger--Robertson inequality, we show that the additional covariance term effectively performs an optimization over the intra-mode rotations entering the criterion. We demonstrate this both for quadrature variables and for number-phase-squeezed states. We further show that Simon's criterion carries out this optimization automatically, which motivates a Simon-like criterion for number-phase-squeezed states that performs the optimization directly in the number-phase plane. We also analyze entanglement in terms of the product of the noises of the two modes, which we call the noise area. Analytically and numerically, we explore whether widely used entanglement criteria can be interpreted as searches for a noise area below unity. In particular, for the product form of the Duan--Giedke--Cirac--Zoller criterion, we show numerically that the minimum noise area obtained after optimization over intra-mode rotations equals the input nonclassicality that a beam splitter needs to generate the same amount of entanglement as the state under consideration. Finally, for Gaussian states we introduce an alternative entanglement measure that can also be extended to multimode settings, and we outline several open questions, including a simpler definition of entanglement depth for number-phase-squeezed states.
\end{abstract}

\maketitle

\section{Introduction}

Quantum optics has evolved from a largely foundational subject into a mature experimental platform for quantum technologies. Quantum cryptography~\cite{BernsteinNatureCryptography2017}, quantum teleportation~\cite{BraunsteinNatureTeleportation2015}, measurements below the standard quantum limit~\cite{MeasurementSQLSciene2004,LIGO2013}, and quantum radars~\cite{QuantumRadarSciRep2017} all rely on genuinely nonclassical resources such as squeezing and entanglement.

Recent advances in plasmonic nano-optics have pushed the generation of nonclassical states toward smaller and less controllable systems, including nanostructures~\cite{bozhevolnyi2017plasmonics,you2020multiparticle,hucPRLk2009squeezedplasmon,tasgin2020wavepackets}. The prospect of integrating such systems into electronic and photonic devices has strengthened the case for miniaturized quantum technologies~\cite{tame2013quantum,Lavrinenko2019IndexEnhancement}. Epsilon-near-zero~(ENZ) materials have also emerged as promising ingredients because they can preserve and transfer coherence over longer distances~\cite{caglayan2017coherence,Lavrinenko2019ENZquantumnetworks}. These developments make reliable detection and characterization of nonclassical states increasingly important.

The nonclassicality used in quantum technologies appears in several forms: single-mode nonclassicality~(SMNc), for example squeezing; two-mode entanglement~(TME); and many-particle entanglement~(MPE). Much like energy in classical mechanics, nonclassicality can be converted from one form to another. A standard example is the conversion of SMNc at the input of a beam splitter~(BS) into TME at its output~\cite{Kim:02,Tahira:09}. Likewise, quadrature squeezing can be transferred into spin squeezing, and hence into MPE, in atomic ensembles~\cite{PolzikPRL1999spinsqz,TothPRA2018SpinSqz,vidal2006concurrence}. TME can also be converted into MPE~\cite{regula2018converting,Tasgin&MeystrePRA2011}.

This convertibility has motivated a broader line of work on how different manifestations of nonclassicality are related. In particular, several studies have examined conservation-like relations between SMNc and TME in beam splitters~\cite{ge2015conservation,arkhipov2016nonclassicality,arkhipov2016interplay,vcernoch2018experimental}, as well as between SMNc and multimode entanglement~\cite{tasgin2020multimode}. Related connections also allow one to derive criteria for SMNc, TME, and MPE from one another~\cite{hillery2006PRA,miranowicz2010testing,gholipour2016entanglement,tasgin2017many,tasgin2020single,tasgin2020measuring}. These relations become especially transparent once one recognizes that quasiparticle excitations over an ensemble become single-mode nonclassical when the ensemble itself is many-particle entangled~\cite{tasgin2017many}.

Despite the remarkable experimental progress in quantum optics and quantum plasmonics, the development of practical nonclassicality tests has been comparatively slow. It remains difficult to obtain experimentally accessible TME measures for states beyond the Gaussian family. A related problem is that most practical entanglement criteria are tailored to particular classes of states.

The most common routes to TME criteria can be summarized as follows. (i) One may construct an inequality that is satisfied by all separable states~\cite{DGCZ_PRL2000,Mancini&TombesiPRL2002_DGCZ_product,RaymerPRA2003,Hillery&ZubairyPRL2006}; its violation then witnesses TME. (ii) One may use the fact that a separable two-mode state remains physical under partial transposition~(PT)~\cite{Peres:96,Horodecki:96}. The PT state must therefore satisfy, for example, the Heisenberg uncertainty relation~(HUR)~\cite{Agarwal&BiswasNJP2005,NhaPRA2006Fock_states} and the Schr\"odinger--Robertson~(SR) inequality~\cite{Nha&ZubairyPRL2008,nha2007entanglement}; violating these PT-based inequalities witnesses TME. (iii) One may further exploit the fact that positive-definite operators must remain positive under PT for separable states~\cite{ShchukinVogelPRL2005,ivan2006generation,ivan2012generation}. (iv) Finally, one may derive new criteria by explicitly relating SMNc, TME, and MPE criteria to one another~\cite{tasgin2017many,hillery2006PRA,tasgin2020single,tasgin2020measuring}. The class of states detected by a given criterion is determined largely by the operators entering the inequality. For example, the Duan--Giedke--Cirac--Zoller~(DGCZ) criterion~\cite{DGCZ_PRL2000}, its product form due to Mancini \textit{et al.}, and the Simon--Peres--Horodecki~(SPH) criterion~\cite{SimonPRL2000} are well suited to quadrature-squeezed-like states, whereas the Hillery--Zubairy~(HZ) criterion~\cite{Hillery&ZubairyPRL2006} is more natural for number-phase-squeezed-like states.

Another route to nonclassicality relies on the negativity, or nonanalyticity, of the Glauber--Sudarshan $P(\alpha_1,\alpha_2)$ function~\cite{ScullyZubairyBook}. One rewrites an operator $\hat{O}$ in normal-ordered form as a positive-definite function. Its expectation value can then be expressed in terms of the $P$ function, schematically as $\langle\hat{O}\rangle=\langle\hat{O}_1\rangle + \int d^2\alpha_1 d^2\alpha_2\, P(\alpha_1,\alpha_2)\,(\ldots)^2$, where $\hat{O}_1$ collects the commutator terms generated during normal ordering. Hence, negativity of $\langle\hat{O}-\hat{O}_1\rangle$ implies negativity of $P(\alpha_1,\alpha_2)$; see Sec.~\ref{sec:Pfnx}. The limitation is equally clear: this approach does not, by itself, distinguish whether the negativity comes from TME, from local SMNc in one or both modes, or from both simultaneously.

The central aim of this paper is therefore not merely to add another criterion, but to clarify the internal mechanism of existing entanglement tests. In particular, we ask how intra-mode rotations enter these criteria, why criteria derived from the SR inequality are stronger than their HUR counterparts, and whether several familiar entanglement criteria can be interpreted as searches for a noise area below unity. Here the noise area means the product of the minimum noises of the two modes and therefore refers only to local single-mode nonclassicalities, not to intermode correlations.

Our main results are the following. First, for Gaussian states we revisit entanglement generation at a beam splitter and give a geometric interpretation in terms of input and output noise areas. This viewpoint naturally motivates an alternative entanglement quantifier for Gaussian states, expressed in the same units as nonclassical depth. Second, for PT-based criteria derived from the SR inequality, we show that the additional covariance term compensates for nonoptimal choices of the mixing directions; in effect, it performs part of the optimization over intra-mode rotations. Third, we show that Simon's criterion performs this optimization automatically. This observation then leads us to a Simon-like criterion for number-phase-squeezed-like states, where the optimization takes place in the $n$--$\Phi$ plane. Fourth, we present analytic arguments and numerical evidence that optimized DGCZ-type criteria behave like searches for a noise area below unity. In particular, the minimum of the product-form DGCZ criterion over intra-mode rotations matches the beam splitter input nonclassicality required to generate the same logarithmic negativity as the state under study. Several parts of the paper are best read as structural observations and motivating connections rather than as general no-go or equivalence theorems.

The paper is organized as follows. Section~\ref{sec:GaussianStates} deals with Gaussian states. We review the covariance-matrix description of SMNc, derive the beam splitter output covariance matrix, introduce the noise area viewpoint, and discuss how it relates to logarithmic negativity and to an alternative entanglement measure. We then show how noise area inequalities emerge from the $P$ function, analyze the role of the extra term in the SR inequality, and present our numerical observations linking optimized DGCZ-type criteria to beam splitter input noise areas. Section~\ref{sec:number_phase_states} extends the discussion to number-phase-squeezed-like states. There we show that the relevant covariance-matrix eigenvalues are invariant under rotations and displacements in the $n$--$\Phi$ plane, analyze the extra term in the SR-based criterion, derive a Simon-like criterion for these states, and discuss the corresponding noise area picture. Section~\ref{sec:summary} summarizes the main conclusions and highlights open questions.


\section{Gaussian states} \label{sec:GaussianStates}

In this section we analyze the nonclassical properties of Gaussian states and introduce the notation used throughout the paper.

\subsection{Properties of nonclassical Gaussian states} \label{sec:NonclassicalityGaussian}

For Gaussian states, all nonclassical features are encoded in the covariance~(noise) matrix
$V_{ij}^{\rm (r)}=\langle \hat{u}_i \hat{u}_j\rangle/2- \langle \hat{u}_i\rangle \langle \hat{u}_j\rangle$~\cite{Braunstein:05},
where the vector is $u=[x_1,p_1,x_2,p_2]$ for a two-mode~(TM) state and $u=[x_1,p_1]$ for a single-mode~(SM) state. The Wigner function of a Gaussian state is completely determined by $V^{\rm (r)}$.

It is often convenient to pass from the real representation to the complex one, $u^{\rm (c)}=[\alpha_1,\alpha_1^\ast,\alpha_2,\alpha_2^\ast]$, using the transformation matrices~\cite{simon1994quantum} $\mathcal{T}_1=[1,i;1,-i]/\sqrt{2}$ and $\mathcal{T}_2=\mathcal{T}_1\otimes\mathcal{T}_1$ for SM and TM states, respectively. Thus, $V^{\rm (c)}=\mathcal{T} V^{\rm (r)}\mathcal{T}^\dagger$, where $\alpha_{1,2}=(x_{1,2}+ip_{1,2})/\sqrt{2}$. The SM case follows in the same way.

For a single-mode Gaussian state, the complex covariance matrix can be written as
\begin{eqnarray}
V^{\rm (c)}=\begin{bmatrix}
    a   & b&  \\
    b^\ast  & a 
\end{bmatrix} \: ,
\label{Vc_gaussian}
\end{eqnarray}
where $a=\langle \hat{a}_1^\dagger \hat{a}_1 \rangle +1/2$ and $b=\langle \hat{a}_1^2 \rangle$. A SM Gaussian state is nonclassical when $|\langle \hat{a}_1^2\rangle|>\langle \hat{a}_1^\dagger \hat{a}_1 \rangle$, that is, when $|b|>a-1/2$. The eigenvalues of $V^{\rm (c)}$, $\Lambda_{\rm lg,sm}=a\pm |b|$, give the largest~(lg) and smallest~(sm) quadrature noises accessible by intra-mode rotations. The state is squeezed below the standard quantum limit~(SQL) when $\Lambda_{\rm sm}<1/2$. For later convenience we define $\lambda=2\Lambda$, so that $\lambda<1$ signals quadrature squeezing, equivalently SMNc.

The role of the rotation becomes explicit if we write $\hat{a}_{\phi_1}=e^{-i \phi_1} \hat{a}_1$ and examine the quadrature $x_{\phi_1}$. One finds
\begin{eqnarray}
\langle (\Delta\hat{x})^2_{\phi_1} \rangle=\langle \hat{a}_1^\dagger \hat{a}_1 \rangle +\frac{1}{2}+ \frac{1}{2}(b e^{-i2\phi_1}+b^\ast e^{i2\phi_1}).
\label{xphi}
\end{eqnarray}
The minimum noise is obtained by choosing $\phi_1$ so that the last term takes the value $-|b|$, whereas the maximum noise corresponds to $+|b|$. The eigenvalues of Eq.~(\ref{Vc_gaussian}) perform this optimization automatically. For a quadrature-squeezed state, $\langle (\Delta x_{\phi_1})^2 \rangle=e^{-2r}/2=a-|b|$ and $\langle (\Delta p_{\phi_1})^2 \rangle=e^{2r}/2=a+|b|$. Throughout the paper we assume, for brevity, that a local unitary transformation has been used to set first moments to zero~\cite{simon1994quantum}; this does not affect the covariance-matrix properties relevant to SMNc.

A more general quantifier of nonclassicality is the nonclassical depth $\tau$~\cite{lee1991measure}. It is defined as the minimum filtering strength in
\begin{equation}
R(\alpha,\tau)=\frac{1}{\pi\tau} \int d^2\alpha' \: \exp\left(-|\alpha-\alpha'|^2/\tau\right) \: P(\alpha') \: , 
\label{tauR}
\end{equation}
that renders the Glauber--Sudarshan function $P(\alpha)$ analytic and positive definite. The Gaussian factor is a nonclassicality filter~\cite{vogelPRA2010Filters}, and $d^2\alpha\equiv d\alpha_R \, d\alpha_I$ denotes integration over the real and imaginary parts of $\alpha$.

For Gaussian states, the nonclassical depth is simply~\cite{lee1991measure}
$\tau={\rm max}\{0,(1-\lambda_{\rm sm})/2\}$. Thus nonclassicality becomes nonzero once the minimum noise drops below the SQL, $\lambda_{\rm sm}<1$.

\subsection{After the beam splitter} \label{sec:afterBS}

Consider two single-mode Gaussian states mixed at a beam splitter. Let the first input have covariance matrix $V_1^{\rm (c)}$, with $V_1^{\rm (c)}=V^{\rm (c)}$ from Eq.~(\ref{Vc_gaussian}), and let the second input be described by
\begin{eqnarray}
V_2^{(c)}=\begin{bmatrix}
    a_2  & b_2&  \\
    b_2^*  & a_2
\end{bmatrix} \: ,
\label{V2matrix}
\end{eqnarray}
Then the two-mode covariance matrix at the beam splitter output is
\begin{eqnarray}
V_{\rm BS}^{\rm (c)}=\begin{bmatrix}
    A  & C&  \\
    C^\dagger & B
\end{bmatrix}
\label{VBS}
\end{eqnarray}
with
\begin{eqnarray}
A= \left( \begin{array}{cc}
a_1 \cos^2\theta_{\scaleto{\rm BS}{3 pt}}+a_2\sin^2\theta_{\scaleto{\rm BS}{3 pt}} & b_1\cos^2\theta_{\scaleto{\rm BS}{3 pt}}+b_2\sin^2\theta_{\scaleto{\rm BS}{3 pt}} \\
 b_1^{\ast}\cos^2\theta_{\scaleto{\rm BS}{3 pt}}+b_2^{\ast}\sin^2\theta_{\scaleto{\rm BS}{3 pt}}& a_1 \cos^2\theta_{\scaleto{\rm BS}{3 pt}}+a_2\sin^2\theta_{\scaleto{\rm BS}{3 pt}}
\end{array} \right),\nonumber\\
\label{Amat}
\end{eqnarray}
\begin{eqnarray}
B= \left( \begin{array}{cc}
a_1 \sin^2\theta_{\scaleto{\rm BS}{3 pt}}+a_2\cos^2\theta_{\scaleto{\rm BS}{3 pt}} & b_1\sin^2\theta_{\scaleto{\rm BS}{3 pt}} +b_2\cos^2\theta_{\scaleto{\rm BS}{3 pt}} \\
 b_1^{\ast}\sin^2\theta_{\scaleto{\rm BS}{3 pt}} +b_2^{\ast}\cos^2\theta_{\scaleto{\rm BS}{3 pt}} & a_1 \sin^2\theta_{\scaleto{\rm BS}{3 pt}}+a_2\cos^2\theta_{\scaleto{\rm BS}{3 pt}}
\end{array} \right),\nonumber\\
\label{Bmat}
\end{eqnarray}
\begin{eqnarray}
C= \sin\theta_{\scaleto{\rm BS}{3 pt}}\cos\theta_{\scaleto{\rm BS}{3 pt}}
\left( \begin{array}{cc}
(a_1-a_2)& (b_1-b_2) \\
 (b_1-b_2)^{\ast} & (a_1-a_2)
\end{array} \right) .\nonumber\\
\label{Cmatrix}
\end{eqnarray}
Here, $\theta_{\scaleto{\rm BS}{3 pt}}$ is the beam splitter mixing angle and $R=r^2=\sin^2\theta_{\scaleto{\rm BS}{3 pt}}$ is the reflection coefficient.

\subsection{Noise area and beam splitter entanglement} \label{sec:NA?}

The TME generated at the beam splitter output can be quantified by the logarithmic negativity~\cite{adesso2004extremal,Vidal:02}, $E_{\mathcal{N}}$. This quantity is closely related to the Simon--Peres--Horodecki~(SPH) criterion~\cite{SimonPRL2000} and is an entanglement monotone~\cite{plenio2005logarithmic}. For the mixing of two nonclassical states, Ref.~\cite{Li:06} finds
\begin{equation}
E_{\mathcal{N}}={\rm max}\{ 0, -\frac{1}{2} \log_2[(1-2\tau_1)(1-2\tau_2)] \},
\label{EN_2006}
\end{equation}
where $\tau_{1,2}$ are the nonclassical depths of the two input modes. Likewise, when a nonclassical state is mixed with a thermal~(noisy) state, Ref.~\cite{Tahira:09} obtains
\begin{equation}
E_{\mathcal{N}}={\rm max}\{ 0, -\frac{1}{2} \log_2[(1-2\tau_1)(1+2\bar{n})]\},
\label{EN_Tahira}
\end{equation}
where $\bar{n}$ is the mean photon number of the thermal state, and $(1+2\bar{n})$ is precisely its minimum noise $\lambda_{\rm sm}$.

Equations~(\ref{EN_2006}) and~(\ref{EN_Tahira}) can therefore be written in the common form
\begin{equation}
E_{\mathcal{N}}={\rm max}\{ 0, -\frac{1}{2} \log_2[\lambda_{\rm sm}^{(1)}\lambda_{\rm sm}^{(2)}] \}.
\label{ENnoise}
\end{equation}
For Gaussian states, entanglement is generated at the BS output whenever the input \emph{noise area} is smaller than unity, namely $\Omega^{\rm (in)}=\lambda_{\rm sm}^{(1)}\lambda_{\rm sm}^{(2)}<1$.

An equally useful quantity is
\begin{equation}
S_{\mathcal{N}}=\log_2 \frac{\lambda_{1,{\rm sm}}^{\rm (out)} \: \lambda_{2,{\rm sm}}^{\rm (out)}}{\lambda_{1,{\rm sm}}^{\rm (in)} \: \lambda_{2,{\rm sm}}^{\rm (in)}},
\label{SN}
\end{equation}
namely Eq.~(11) of Ref.~\cite{ge2015conservation}. This quantity behaves as an entanglement measure equivalent to the logarithmic negativity for the beam splitter output~\cite{ge2015conservation,arkhipov2016nonclassicality,arkhipov2016interplay}. Equation~(\ref{SN}) shows that the generated entanglement grows with the ratio of output to input noise area. Here $\lambda_{1,2\:{\rm sm}}^{\rm (out)}$ denote the SMNc that remains in the two output modes after the BS has converted part of the initial local nonclassicality into TME~\cite{ge2015conservation}. These remaining local noises are obtained by deleting the correlation block $C$ from Eqs.~(\ref{VBS}) and~(\ref{Cmatrix}). An analogous behavior also appears for multimode entanglement~\cite{tasgin2020multimode}.

These observations motivate the main question of this section: does the noise area play a deeper role in the structure of TME criteria~\cite{DGCZ_PRL2000,Mancini&TombesiPRL2002_DGCZ_product,RaymerPRA2003,Hillery&ZubairyPRL2006,Agarwal&BiswasNJP2005,Nha&ZubairyPRL2008,NhaPRA2006Fock_states}? To explore this possibility, we now reconsider beam splitter mixing from the noise area viewpoint.

\subsubsection{Beam splitter output noise area} \label{sec:outputNA}

The unconverted SMNc remaining in the two BS output modes can be isolated as follows~\cite{ge2015conservation}. If we erase the correlations by setting the $2\times2$ block $C=[0,0;0,0]$ in Eq.~(\ref{VBS}), the only nonclassicality left in the state is the local SMNc in the two output modes. The corresponding minimum noises are
\begin{eqnarray}
\lambda_{1,{\rm min}}^{\rm (out)}=(a_1-|b_1|) \cos^2\theta_{\scaleto{\rm BS}{3 pt}} + (a_2-|b_2|) \sin^2\theta_{\scaleto{\rm BS}{3 pt}},  \qquad
\label{lambda1}
\\
\lambda_{2,{\rm min}}^{\rm (out)}=(a_2-|b_2|) \cos^2\theta_{\scaleto{\rm BS}{3 pt}} + (a_1-|b_1|) \sin^2\theta_{\scaleto{\rm BS}{3 pt}}, \qquad
\label{lambda2}
\end{eqnarray}
which give the increased output noise area
\begin{eqnarray}
\Omega^{\rm (out)} = &&(a_1-|b_1|)((a_2-|b_2|)) \\
+&& \sin^2\theta_{\scaleto{\rm BS}{3 pt}}\cos^2\theta_{\scaleto{\rm BS}{3 pt}} \left[ (a_1-|b_1|) - (a_2-|b_2|) \right]^2 \qquad
\label{NA_out}
\end{eqnarray}
compared with the initial noise area
\begin{eqnarray}
\Omega^{\rm (in)} = (a_1-|b_1|)(a_2-|b_2|),
\label{NA_in}
\end{eqnarray}
where the two input modes are separable.

\begin{figure}
\includegraphics[trim=0cm 0cm 0cm 0cm, clip, width=0.4\textwidth]{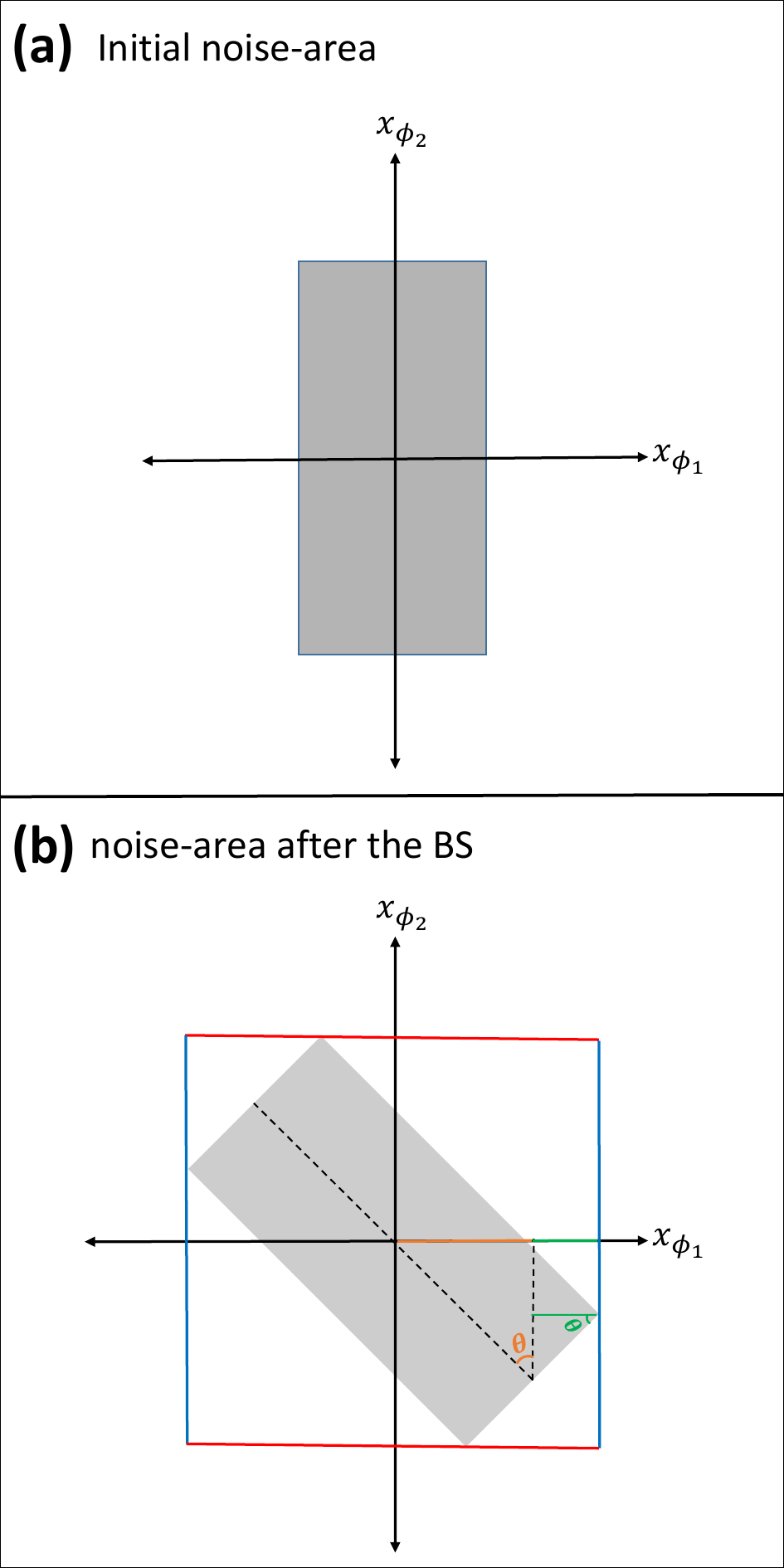}
\caption{(Color online) Geometric illustration of the conversion of local nonclassicality into TME~($\tau_{\rm ent}$). The shaded rectangle represents the initial local nonclassicality, while the outer rectangle defined by the red and blue lines represents the increased output noise area. The directions $x_{\phi_1}$ and $x_{\phi_2}$ are chosen along the minimum-noise quadratures of the two input modes. See the text for details.}
\label{fig:geometry}
\end{figure}

Figure~\ref{fig:geometry} illustrates the coordinate transformation induced by the BS, where $\theta\equiv\theta_{\rm \scriptscriptstyle BS}$. The green and orange lines in Fig.~\ref{fig:geometry}(b) correspond to the two terms entering Eq.~(\ref{lambda1}); Eq.~(\ref{lambda2}) is interpreted in the same way. Strictly speaking, the area of the outer rectangle in Fig.~\ref{fig:geometry}(b) is only a geometric illustration of $\Omega^{\rm (out)}$, because Eqs.~(\ref{lambda1}) and~(\ref{lambda2}) involve sums of squared transformed coordinates. Even so, the figure captures the basic idea well: the noise area grows as local SMNc is converted into TME by the BS.

The ratio of output to input noise area is
\begin{eqnarray}
\frac{\Omega^{\rm (out)} }{\Omega^{\rm (in)} }&=& 1 + \frac{(\lambda_{1,{\rm sm}}-\lambda_{2,{\rm sm}})^2}{\lambda_{1,{\rm sm}}\lambda_{2,{\rm sm}}} \nonumber \\
&=& 1 + \sin^2\theta_{\rm \scriptscriptstyle BS} \cos^2\theta_{\rm \scriptscriptstyle BS} \frac{(a_1-a_2 + |b_2|-|b_1|)^2}{(a_1-|b_1|)(a_2-|b_2|)}, \qquad \quad
\label{NAratio}
\end{eqnarray}
whose inverse is proportional to the logarithmic negativity of the output state~\cite{ge2015conservation}. The increase in noise area is maximal at $\theta_{\scaleto{\rm BS}{3 pt}}=\pi/4$, where the two output noises become equal. This is precisely the optimal mixing angle for an ideal lossless BS. It is also immediate from Eq.~(\ref{NA_out}) that the noise area does not change when the two input modes have the same minimum noise. In that case, as expected, an ideal BS does not generate entanglement.

\subsubsection{Alternative entanglement measure} \label{sec:alternativeTME}

In Ref.~\cite{ge2015conservation} and in Sec.~\ref{sec:outputNA}, we isolated the unconverted SMNc in the BS output by deleting the correlation block $C$ from the covariance matrix~(\ref{VBS}). The complementary construction is to remove the local SMNc instead and keep only the nonclassicality that originates from entanglement. To do so, we replace the local blocks $A$ and $B$ by vacuum blocks, $I/2$, and define
\begin{eqnarray}
V_{\rm ent}^{(c)}=\begin{bmatrix}
    I/2  & C&  \\
    C^\dagger & I/2
\end{bmatrix},
\label{Vent}
\end{eqnarray}
so that the remaining nonclassicality is due solely to entanglement. Here $I$ is the $2\times2$ identity matrix, whose nonclassical depth is that of a coherent or vacuum state.

The nonclassicality of $V_{\rm ent}^{(c)}$ can be determined from the condition ${\rm eig}[V_{\rm ent}^{(c)}+{\cal\tau}]>0$~\cite{Li:06}. If one constrains the filter to inject the same amount of noise into both modes, $\tau={\rm diag}[\tau_1,\tau_1,\tau_1,\tau_1]$, one obtains the minimum common noise needed to render the corresponding $P$ function positive. However, that constraint can in general force one to add more noise than necessary~\footnote{\label{fn:tautau2} In Ref.~\cite{Li:06}, the same filtering strength $\tau$ is imposed on both modes. More generally, one may allow different strengths, so that the Gaussian filter injects $\tau_1$ and $\tau_2$ into the two modes separately. This can reduce the total injected noise~\cite{tasgin2020multimode}, since one may have $\tau^2>\tau_1\tau_2$, or equivalently $(1-2\tau)^2<(1-2\tau_1)(1-2\tau_2)$.}. For that reason, here and in Ref.~\cite{tasgin2020multimode}, we use the more general filter matrix $\tau={\rm diag}[\tau_1,\tau_1,\tau_2,\tau_2]$. The eigenvalue condition becomes
\begin{eqnarray}
{\rm eig}\begin{bmatrix}
\frac{1}{2}+\tau_1 &  0  &   \tilde{a}  &  \tilde{b}
\\
0 & \frac{1}{2}+\tau_1   &  \tilde{b}^* & \tilde{a}
\\
\tilde{a} & \tilde{b} & \frac{1}{2}+\tau_2   & 0
\\
\tilde{b}^* & \tilde{a}  & 0  & \frac{1}{2}+\tau_2
\end{bmatrix} =0,
\label{eigVent}
\end{eqnarray}
where $\tilde{a}=\sin\theta_{\scaleto{\rm BS}{3 pt}}\cos\theta_{\scaleto{\rm BS}{3 pt}}(a_1-a_2)$ and $\tilde{b}=\sin\theta_{\scaleto{\rm BS}{3 pt}}\cos\theta_{\scaleto{\rm BS}{3 pt}}(b_1-b_2)$. The equation for the eigenvalues $\beta_{1-4}$ can be written as
\begin{eqnarray}
\tilde{\tau}_1^2\tilde{\tau}_2^2 - (\tilde{a}^2+|\tilde{b}|^2)\tilde{\tau}_1\tilde{\tau}_2 + (\tilde{a}^2-|\tilde{b}|^2)^2=0,
\end{eqnarray}
where $\tilde{\tau}_{1,2}=1/2+\tau_{1,2} - \beta$. Defining $x=\tilde{\tau}_1\tilde{\tau}_2$, one obtains
\begin{equation}
x^{(\pm)}=\tilde{\tau}_1\tilde{\tau}_2 = (|\tilde{a}|\pm |\tilde{b}|)^2,
\end{equation}
so positivity of the eigenvalues requires
\begin{equation}
(1/2+\tau_1)(1/2+\tau_2) > (|\tilde{a}|\pm |\tilde{b}|)^2.
\end{equation}
To make all eigenvalues positive, it is enough to impose
\begin{equation}
(1/2+\tau_1)(1/2+\tau_2) > (|\tilde{a}| + |\tilde{b}|)^2.
\label{tau1tau2}
\end{equation}

For the present beam splitter problem, the minimum injected noise occurs at $\tau_1=\tau_2$. One then obtains $\tau_1={\rm max}[0,(|\tilde{a}|+|\tilde{b}|-1/2)]$. The noise area associated with the entanglement in $V_{\rm ent}^{(c)}$ is therefore
\begin{equation}
\Omega_{\rm ent}=(1-2\tau_1)(1-2\tau_2).
\label{OmegaEnt}
\end{equation}
(See Ref.~\cite{tasgin2020multimode} for the multimode generalization.) Smaller values of $\Omega_{\rm ent}$ correspond to stronger entanglement, because they mean that more noise must be injected to eliminate the nonclassicality of $V_{\rm ent}^{(c)}$.

Thus the entanglement contained in the BS state $V_{\rm BS}^{\rm (c)}$, Eq.~(\ref{VBS}), can be expressed in the same noise area units as the local SMNc,
\begin{equation}
\Omega_{\rm ent}= \left[  1- \sin\theta_{\scaleto{\rm BS}{3 pt}}\cos\theta_{\scaleto{\rm BS}{3 pt}}(|a_1-a_2|+|b_1-b_2|)  \right]^2.
\label{AreaEnt}
\end{equation}
This is useful because it places entanglement and residual local nonclassicality, such as $\Omega^{\rm (out)}$ in Eq.~(\ref{NA_out}), on the same scale.

One may also write the entanglement contribution in terms of the nonclassical depth
\begin{equation}
\tau_{\rm ent}=|\sin\theta_{\scaleto{BS}{3 pt}} \cos\theta_{\scaleto{BS}{3 pt}}| [|a_1-a_2|+|b_1-b_2|]
\label{tauent}
\end{equation}
with the caveat that the noise $\tau_{\rm ent}$ is injected into \emph{both} modes. Quantitatively, the corresponding reduction in noise therefore appears more naturally as $(1-2\tau_{\rm ent})^2$. In the simple BS setting considered here, the optimal injection happens to be symmetric, $\tau_1=\tau_2$. In more general multimode settings, the optimal filtering need not be symmetric~\cite{tasgin2020multimode}.

\par\noindent\rule{0.45 \textwidth}{0.4pt}

A useful consequence of this construction is that $\Omega_{\rm ent}$ decreases as $\Omega^{\rm (out)}/\Omega^{\rm (in)}$ increases. In other words, as more local SMNc is converted into entanglement, the remaining local noise area grows while the entanglement-associated noise area shrinks. Although the product $\big(\Omega^{\rm (out)}/\Omega^{\rm (in)}\big)\Omega_{\rm ent}$ is not constant, the exchange between the two quantities makes the conversion of local nonclassicality into TME directly visible. This provides an alternative viewpoint on the conservation-like relation derived in Ref.~\cite{ge2015conservation} and related works~\cite{arkhipov2016nonclassicality,arkhipov2016interplay}.

It is also worth stressing that this picture is specific to BS-generated entanglement. For example, a two-mode squeezing Hamiltonian $\exp( \xi\hat{a}_1^\dagger\hat{a}_2^\dagger - \xi^*\hat{a}_1 \hat{a}_2)$ generates pure entanglement from coherent-state inputs, and in that case the residual local SMNc obtained by setting $C=0$ in Eq.~(\ref{VBS}) vanishes.

\subsubsection{Do entanglement criteria search for a noise area below unity?} \label{sec:NAbelow1}

The observation in Fig.~\ref{fig:geometry}, together with Eq.~(\ref{ENnoise}), suggests a natural inverse question. If we conceptually reverse the beam splitter process, Fig.~\ref{fig:geometry}(b)$\to$Fig.~\ref{fig:geometry}(a), do familiar entanglement criteria effectively rotate a two-mode state backward and test whether the resulting noise area falls below unity? More concretely, do they seek a configuration for which $\langle(\Delta\hat{x}_{\phi_1})^2\rangle \, \langle(\Delta\hat{x}_{\phi_2})^2\rangle < 1/4$, or equivalently $\lambda_{\phi_1}\lambda_{\phi_2}<1$?

Not every entangled two-mode state can itself be produced by a single BS. Nevertheless, any two-mode state can be back-rotated by an angle $-\theta_{\scaleto{\rm BS}{3 pt}}$, and the resulting state can be examined for a sufficiently small product of local minimum noises. Importantly, the quantities $\langle(\Delta\hat{x}_{\phi_{1,2}})^2\rangle$ entering this test refer only to local single-mode noises; they do not involve intermode correlations. As we show next, a noise area below unity is itself a nonclassicality condition and is therefore a necessary ingredient for BS generation of TME~\cite{Kim:02}.

The directions $\hat{x}_{\phi_{1,2}}$ in Fig.~\ref{fig:geometry}(a) are chosen along the minimum-noise axes of the two modes. Any other choice would give a larger value of the noise area. Once again, the eigenvalue $\lambda_{\rm sm}$ performs this optimization automatically. Simon's criterion~\cite{SimonPRL2000}, and likewise the logarithmic negativity $E_{\mathcal{N}}$, share this crucial feature because they are invariant under local rotations $\hat{a}_{1,2}(\phi_{1,2})=e^{i\phi_{1,2}}\hat{a}_{1,2}$. The DGCZ criterion~\cite{DGCZ_PRL2000} and its product form~\cite{Mancini&TombesiPRL2002_DGCZ_product,RaymerPRA2003}, by contrast, are not invariant under such rotations. If one wants the strongest DGCZ-type witness, one must therefore optimize over the local directions explicitly.

Our purpose here is not to claim a fully general equivalence between TME criteria and a BS search for a noise area below unity. Rather, we aim to isolate and clarify a structural pattern that repeatedly appears in well-known criteria.

\subsubsection{Noise area as a nonclassicality condition} \label{sec:Pfnx}

We now show that a noise area below unity is itself a nonclassicality condition: it implies that the Glauber--Sudarshan $P$ function is negative or nonanalytic in some region. Crucially, the resulting inequalities do \emph{not} refer to intermode correlations; they probe only the local noises of the two modes~\footnote{We recall that $P(\alpha_1,\alpha_2)<0$ is a necessary and sufficient condition for TME generation at the BS output~\cite{Kim:02}.}.

As a representative example, consider the operator identity
\begin{eqnarray}
\langle(\hat{n}_1-&&\hat{n}_2)^2\rangle  = \langle\hat{n}_1+\hat{n}_2\rangle \hspace{5 cm} \nonumber \\
\\
&&+  \langle  \hat{a}_1^\dagger\hat{a}_1^\dagger\hat{a}_1 \hat{a}_1 + \hat{a}_2^\dagger\hat{a}_2^\dagger\hat{a}_2 \hat{a}_2
-2\hat{a}_1^\dagger\hat{a}_2 \hat{a}_2 \hat{a}_1^\dagger\hat{a}_1 \hat{a}_1  \rangle,
\label{NO_n1minusn2}
\end{eqnarray}
where the last term is in normal order and can therefore be written in the $P$ representation as
\begin{eqnarray}
\langle(\hat{n}_1-&&\hat{n}_2)^2\rangle = \langle\hat{n}_1+\hat{n}_2\rangle \hspace{5 cm} \nonumber \\
\\
&&+ \int d^2\alpha_1d^2\alpha_2 \: P(\alpha_1,\alpha_2) \: (|\alpha_1|^2-|\alpha_2|^2)^2.
\label{NcHillery}
\end{eqnarray}
Hence, if $\langle(\hat{n}_1-\hat{n}_2)^2\rangle < \langle\hat{n}_1+\hat{n}_2\rangle$, the only possible origin is negativity of $P(\alpha_1,\alpha_2)$ in some finite region, which proves that $\hat{\rho}$ is nonclassical. The term $\langle \hat{n}_1 + \hat{n}_2\rangle$ appears because of the commutators generated during normal ordering; this point will be important again in Sec.~\ref{sec:ObservNA}.

The same logic yields nonclassicality conditions that depend only on local noises. For quadratures one finds
\begin{eqnarray}
\langle(\Delta\hat{x}_1)^2\rangle \langle(\Delta\hat{x}_2)^2\rangle < 1/4 \quad \text{or},
\label{Ncx1x2}
\\
\lambda_{1,{\rm sm}} \lambda_{2,{\rm sm}} < 1,
\label{Nclamlam2}
\end{eqnarray}
provided that $\hat{x}_{1,2}$ are chosen along the minimum-noise directions. A weaker form is
\begin{eqnarray}
\langle(\Delta\hat{x}_1)^2\rangle + \langle(\Delta\hat{x}_2)^2\rangle < 1
\label{Ncx1sumx2}
\end{eqnarray}
which parallels the relation between the DGCZ sum criterion and the stronger product criterion of Mancini \textit{et al.}~\cite{Mancini&TombesiPRL2002_DGCZ_product}.

For number-squeezed-like states one similarly obtains
\begin{equation}
\Omega_n=\frac{\langle(\Delta \hat{n}_1)^2\rangle}{\langle\hat{n}_1\rangle} \: \frac{\langle(\Delta \hat{n}_2)^2\rangle}{\langle\hat{n}_2\rangle} <1
\label{Ncn1n2}
\end{equation}
and the weaker form
\begin{equation}
\langle(\Delta \hat{n}_1)^2\rangle + \langle(\Delta \hat{n}_2)^2\rangle < \langle\hat{n}_1+\hat{n}_2\rangle.
\end{equation}
Since $\langle(\Delta \hat{n})^2\rangle/\langle\hat{n}\rangle<1$ is number squeezing below the SQL, it is natural to refer to the left-hand side of Eq.~(\ref{Ncn1n2}) as the noise area $\Omega_n$ for number-squeezed-like states.

To emphasize the main point once more: these inequalities are two-mode nonclassicality conditions built purely from local single-mode noises. They do not involve intermode correlations explicitly.

\subsection{Role of the extra term in the Schr\"odinger--Robertson inequality} \label{sec:extratermGaussian}

We now explain why criteria obtained from the Schr\"odinger--Robertson~(SR) inequality~\cite{nha2007entanglement,Nha&ZubairyPRL2008} are stronger than those obtained from the Heisenberg uncertainty relation. The key point is that the additional covariance term in the SR inequality compensates, at least partly, for a nonoptimal choice of local quadrature directions.

DGCZ~\cite{DGCZ_PRL2000} and Mancini \textit{et al.}~\cite{Mancini&TombesiPRL2002_DGCZ_product} construct inseparability criteria from the sum or product of the noises of two collective observables. The same criteria can also be derived by imposing the HUR on the PT of a separable state~\cite{Agarwal&BiswasNJP2005}. Later work~\cite{Nha&ZubairyPRL2008,nha2007entanglement,NhaPRA2006Fock_states} showed that a stronger witness follows by using the SR inequality
\begin{equation}
\langle (\Delta\tilde{u})^2\rangle \: \langle (\Delta\tilde{v})^2\rangle
\geq \frac{1}{4} |\langle[\tilde{u},\tilde{v}]\rangle|^2 + 
\langle \Delta\tilde{u}\Delta\tilde{v}\rangle_{\rm S}^2 \: ,
\label{SR_ineq}
\end{equation}
where $\langle \Delta\tilde{u}\Delta\tilde{v}\rangle_{\rm S} = \frac{1}{2}(\langle \Delta\tilde{u}\Delta\tilde{v} \rangle + \langle \Delta\tilde{v}\Delta\tilde{u} \rangle )$. For the collective variables
\begin{eqnarray}
\tilde{u}=\cos\theta \hat{x}_1 + \sin\theta \hat{x}_2,
\label{uDGCZ}
\\
\tilde{v}=\cos\theta \hat{p}_1 - \sin\theta \hat{p}_2,
\label{vDGCZ}
\end{eqnarray}
the HUR version reproduces the product-form DGCZ criterion, whereas the SR version is stronger because of the extra covariance term.

The physical meaning of that extra term is simple. The DGCZ-type mixing in Eqs.~(\ref{uDGCZ}) and~(\ref{vDGCZ}) uses the coordinates $\hat{x}_{1,2}$ and $\hat{p}_{1,2}$, which need not coincide with the local minimum-noise directions $\hat{x}_{\phi_1}$ and $\hat{x}_{\phi_2}$. If the collective variables are built from the optimal local directions, the extra term vanishes. If not, the unused local squeezing reappears as a nonzero value of $\langle \Delta\tilde{u}\Delta\tilde{v}\rangle_{\rm S}$, making the SR witness stronger than its HUR counterpart.

\begin{figure}
\includegraphics[width=0.5\textwidth]{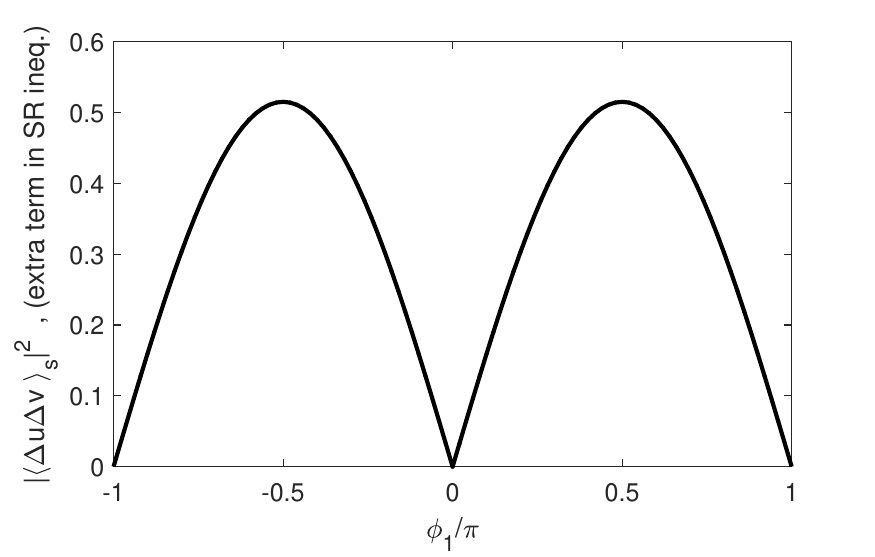}
\caption{Role of the extra term in the entanglement criterion obtained from the SR inequality~(\ref{SR_ineq}). When the quadratures entering Eqs.~(\ref{uDGCZ}) and~(\ref{vDGCZ}) are aligned with the local minimum-noise directions, $\phi_1=0$, the extra term vanishes. As $\hat{x}_1$ is rotated away from that direction, the extra term increases and compensates for the inefficient use of local squeezing in mode~1.}
\label{fig:extra_termGaus}
\end{figure}

Figure~\ref{fig:extra_termGaus} illustrates this point numerically. We keep $\hat{x}_2$ aligned with the minimum-noise direction and rotate $\hat{x}_1$ away from its optimum by an angle $\phi_1$. For $\phi_1=0$, the mixing uses the minimum-uncertainty quadratures and the extra term is zero. As the mixing becomes less efficient, the extra term grows and compensates for the local squeezing that the collective variables fail to exploit directly.

This interpretation also clarifies why Simon's criterion should be stronger for Gaussian states. It is well known, and has been shown analytically~\cite{Marian_2018_JPhysA}, that the optimized DGCZ criterion becomes equivalent to the Simon criterion~\footnote{Ref.~\cite{Marian_2018_JPhysA} shows explicitly that the optimized DGCZ criterion~\cite{DGCZ_PRL2000} reduces to Simon's criterion~\cite{SimonPRL2000}.}. The reason is that the SPH criterion~\cite{SimonPRL2000}, and likewise the logarithmic negativity, is invariant under local canonical transformations, including intra-mode rotations. In that sense, it performs the local optimization automatically. Simon's treatment also inherits the SR structure and is invariant under mirror reflection, so it handles both sign choices in the collective momenta on equal footing.

\subsection{Noise area and entanglement} \label{sec:ObservNA}

Sections~\ref{sec:outputNA}, \ref{sec:alternativeTME}, and \ref{sec:NAbelow1}, together with Fig.~\ref{fig:geometry}, show that beam-splitter-generated entanglement is accompanied by an increase in the output noise area and a decrease in the available local SMNc. We now ask the inverse question: given an entangled Gaussian state, can one back-rotate it and recover a local noise area $\Omega=\lambda_{\phi_1,{\rm sm}}\lambda_{\phi_2,{\rm sm}}$ below unity~\footnote{\label{fn:anystateBS} We are aware that a generic two-mode entangled state need not itself be producible by a single BS. Nevertheless, any such state can be back-rotated by an angle $-\theta_{\scaleto{\rm BS}{3 pt}}$, and the rotated state can then be tested for a sufficiently small product of local minimum noises.}~\footnote{\label{fn:notrestricted} Our discussion is not restricted to states generated by a BS. For example, the state may also be produced by a two-mode-squeezing Hamiltonian.}? Since the noise area depends only on the local minimum noises, this question asks whether part of the entanglement can be reinterpreted as a hidden supply of local nonclassicality after an appropriate back-rotation.

Our goal here is not a formal proof of a universal equivalence. Rather, we present a set of analytic observations and numerical results that strongly suggest such a connection. Similar considerations will later reappear for number-phase-squeezed-like states in Sec.~\ref{sec:noisearea_nPhi}.

The numerical procedure is straightforward. For a given entangled Gaussian state~${}^{\ref{fn:notrestricted}}$, we minimize
\begin{equation}
\Omega_{\scaleto{\rm DGCZ}{4 pt}} = \langle(\Delta \tilde{u})^2\rangle \langle(\Delta \tilde{v})^2\rangle,
\label{Omega_dgcz}
\end{equation}
namely the product-form DGCZ quantity, over the local rotations $\hat{a}_{\phi_{1,2}}=e^{i\phi_{1,2}}\hat{a}_{1,2}$. The collective variables $\tilde{u}$ and $\tilde{v}$ are given by Eqs.~(\ref{uDGCZ}) and~(\ref{vDGCZ}). For the same state we also compute the logarithmic negativity $E_\mathcal{N}$.

The resulting observations are as follows. \emph{(i)} If we define $\Omega_{\rm neg}$ by inverting Eq.~(\ref{ENnoise}), namely $E_\mathcal{N}=-\frac{1}{2}\log_2\Omega_{\rm neg}$, then numerically we find
$\Omega_{\rm neg}=\Omega_{\scaleto{\rm DGCZ}{4 pt}}^{\rm (min)}$.
Equivalently, the amount of entanglement already present in the state is exactly the amount that a BS would generate from an input local noise area equal to the optimized DGCZ product.

This observation is striking because the two quantities are conceptually very different. The quantity $\Omega_{\rm neg}=2^{-2E_{\cal N}}$ is inferred from a genuine two-mode entanglement measure, whereas $\Omega_{\scaleto{\rm DGCZ}{4 pt}}^{\rm (min)}=\left[ (\Delta\tilde{u})^2 (\Delta\tilde{v})^2 \right]^{\rm min}_{\phi_1,\phi_2}$ depends only on optimized local noises of the same state. Yet the two agree numerically.

\emph{(ii)} After optimizing over $\phi_{1,2}$, the quantity $\Omega_{\scaleto{\rm DGCZ}{4 pt}}^{\rm (min)}$ is minimized at $\theta=\pm\pi/4$. This remains true even when the entangled state was produced at some different beam splitter angle $\theta_{\scaleto{\rm BS}{3 pt}}\neq \pi/4$ from an initially separable input. This strongly suggests that the optimized DGCZ product is tied to the entanglement itself rather than to the residual SMNc that remains in the output modes~\cite{ge2015conservation}.

\emph{(iii)} For $\theta=\pi/4$, Eqs.~(\ref{uDGCZ}), (\ref{vDGCZ}), and~(\ref{Omega_dgcz}) can be interpreted as a backward BS rotation applied to the entangled state. The criterion then checks whether
\begin{eqnarray}
(\Delta\tilde{u})^2 (\Delta\tilde{v})^2 < 1/4
\label{DutimesDv1o4}
\\
\text{or} \quad (\Delta\tilde{u})^2 (\Delta\tilde{v})^2 < 1.
\label{DusumDv1}
\end{eqnarray}
That is, it checks whether the back-rotated state has a noise area below unity. Equation~(\ref{DusumDv1}) is simply the weaker sum-type counterpart of Eq.~(\ref{DutimesDv1o4}), in the same sense that the original DGCZ criterion is weaker than its product-form version.

\emph{(iv)} For general $\theta\neq\pi/4$, the connection is less obvious because a beam splitter naturally produces
\begin{eqnarray}
u'=\cos\theta \hat{x}_1 + \sin\theta \hat{x}_2
\label{uprime}
\\
v'=\cos\theta \hat{p}_1 + \sin\theta \hat{p}_2,
\label{vprime}
\end{eqnarray}
whereas the DGCZ criterion uses $\tilde{v}$ from Eq.~(\ref{vDGCZ}). In fact, $(\Delta u')^2 + (\Delta v')^2<1$ cannot hold at all, because the HUR imposes $(\Delta u')^2 + (\Delta v')^2\geq \cos^2\theta+\sin^2\theta=1$.

The crucial observation is that the noise area is unchanged under the coordinate transformation $\hat{p}_2 \to -\hat{p}_2$. We stress that this is \emph{not} a partial transposition of the state; it is only a change of coordinates for the same physical state, taking Eq.~(\ref{vprime}) into Eq.~(\ref{vDGCZ}). In the transformed coordinates, the HUR reads $(\Delta u')^2 + (\Delta \tilde{v})^2 \geq |\cos^2\theta-\sin^2\theta|$, and the lower bound vanishes at $\theta=\pi/4$.

\emph{(v)} This invariance under $p_2\to -p_2$ is not accidental. Nonclassicality conditions of the type discussed in Sec.~\ref{sec:Pfnx} are derived by writing normal-ordered operators as positive-definite functions in the $P$ representation. Under $p_2\to -p_2$, one has
$\hat{a}_2^{n_2}{}^\dagger \hat{a}_2^{m_2} \hat{a}_1^{n_1}{}^\dagger \hat{a}_1^{m_1}
\to \hat{a}_2^{m_2}{}^\dagger \hat{a}_2^{n_2} \hat{a}_1^{n_1}{}^\dagger \hat{a}_1^{m_1}$~\cite{Nha&ZubairyPRL2008},
which preserves normal ordering. The positive-definite function therefore remains positive definite, up to complex conjugation of the $\alpha_2$ contribution. Consequently, the transformed inequality is still a valid nonclassicality condition. In this sense, the DGCZ criterion can be viewed as a beam splitter-like search for a back-rotated noise area below unity.

This use of $p_2\to -p_2$ should not be confused with the familiar PT argument for separability. In PT-based entanglement criteria one exploits the fact that a separable state remains physical under partial transposition~\cite{Agarwal&BiswasNJP2005,Nha&ZubairyPRL2008}. Here the same algebraic transformation appears for a different reason: it preserves the normal-ordered structure of a nonclassicality inequality. The commutator terms that arise in this construction, such as $\langle\hat{n}_1+\hat{n}_2\rangle$ in Eq.~(\ref{NcHillery}), may resemble HUR- or SR-type terms, but the physical logic is distinct.

\subsection{Section summary} \label{sec:SummaryGaussian}

For Gaussian states, the covariance-matrix eigenvalues provide the local minimum noises and hence a natural definition of the noise area. We showed that this noise area increases when a beam splitter converts local SMNc into TME, and we introduced an alternative entanglement quantifier, $\tau_{\rm ent}$, that expresses entanglement in the same units as nonclassical depth. We then argued that the extra term in the SR-based criterion compensates for nonoptimal local quadrature choices, whereas Simon's criterion performs the corresponding optimization automatically. Finally, we presented analytic arguments and numerical evidence that DGCZ-type criteria can be interpreted as searches for a back-rotated noise area below unity, with the optimized product form reproducing the beam splitter input nonclassicality associated with the observed logarithmic negativity.

\section{Number-phase-squeezed-like states} \label{sec:number_phase_states}
In this section we extend the discussion to number-phase-squeezed-like states.

\subsection{Single-mode nonclassicality} \label{sec:SMNcnumber}

In Sec.~\ref{sec:NonclassicalityGaussian} we saw that, for Gaussian states, the smallest and largest noises are obtained directly from the eigenvalues of the covariance matrix $V^{\rm (c)}$ in Eq.~(\ref{Vc_gaussian}). Those eigenvalues are already optimized over local quadrature rotations $\hat{a}_\phi=e^{i\phi}\hat{a}$. We now develop the analogous construction for number-phase-squeezed-like states.

We define the annihilation and creation operators
\begin{equation}
\hat{A}_n = \hat{n}+i\gamma\hat{\Phi} \quad \text{and} \quad \hat{A}_n^\dagger = \hat{n}-i\gamma\hat{\Phi},
\label{An}
\end{equation}
where $\hat{\Phi}$ is the phase operator and $\gamma=2\langle\hat{n}\rangle$~\cite{PeggJModOpt1990Intelligent,KitagawaPRA1986numberphase}. The eigenstates of
\begin{equation}
\hat{E}_n=\hat{n} + i\gamma'\hat{\Phi},
\label{Enoper}
\end{equation}
with the generalized coefficient $\gamma'=r2\langle\hat{n}\rangle=r\gamma$, are intelligent states~\cite{PeggJModOpt1990Intelligent}. These states saturate the corresponding uncertainty relation. For $r=1$, one has $\gamma'=\gamma$ and $\hat{E}_n=\hat{A}_n$; in the regime $\langle\hat{n}\rangle \gg 1$, the coherent-state SQL is then set by $\langle(\Delta\hat{n})^2\rangle=\langle\hat{n}\rangle$ and $\langle(\Delta\hat{\Phi})^2\rangle=1/4\langle\hat{n}\rangle$~\cite{KitagawaPRA1986numberphase}. The eigenstates of $\hat{E}_n$ exhibit number squeezing below the SQL, $\langle(\Delta\hat{n})^2\rangle<\langle\hat{n}\rangle$, for $r<1$, and phase squeezing below the SQL for $r>1$. In both cases the uncertainty product remains minimal, $\langle(\Delta\hat{n})^2\rangle\,\langle(\Delta\hat{\Phi})^2\rangle=1/4$.

The operator in Eq.~(\ref{An}) can also be written in the scaled form
\begin{equation}
\hat{a}_n=\frac{1}{\sqrt{2}} \left( \hat{n} + i\gamma\hat{\Phi}  \right)/\sqrt{\gamma}.
\label{an}
\end{equation}
In analogy with Sec.~\ref{sec:NonclassicalityGaussian}, we define the real covariance matrix
\begin{eqnarray}
\resizebox{0.48 \textwidth}{!}
{$
V_n^{\rm (r)}=\begin{bmatrix}
    \langle(\Delta\hat{n}')^2\rangle  & \langle \hat{n}'\hat{\Phi}' + \hat{\Phi}'\hat{n}'\rangle/2 - \langle\hat{n}'\rangle\langle\hat{\Phi}'\rangle&  \\
    \langle \hat{n}'\hat{\Phi}' + \hat{\Phi}'\hat{n}'\rangle/2 - \langle\hat{n}'\rangle\langle\hat{\Phi}'\rangle  &  \langle(\Delta\hat{\Phi}')^2\rangle 
\end{bmatrix} $} \nonumber
\\
\label{Vr_number}
\end{eqnarray}
in the real representation $\hat{\xi}=[\hat{n}',\hat{\Phi}' ]$, where $\hat{n}'=\hat{n}/\gamma$ and $\hat{\Phi}'=\sqrt{\gamma} \, \hat{\Phi}$ are scaled variables. The corresponding complex covariance matrix is
\begin{eqnarray}
V_n^{\rm (c)}=\begin{bmatrix}
    \langle\hat{a}_n^\dagger\hat{a}_n\rangle - |\langle\hat{a}_n\rangle|^2 + 1/2 &  (\langle\hat{a}_n^2\rangle-\langle\hat{a}_n\rangle^2) &  \\
     (\langle\hat{a}_n^2\rangle-\langle\hat{a}_n\rangle^2)^* & \langle\hat{a}_n^\dagger\hat{a}_n\rangle - |\langle\hat{a}_n\rangle|^2 + 1/2
\end{bmatrix} \nonumber
\\
\label{Vc_number}
\end{eqnarray}
in the representation $\xi=[\alpha_n,\alpha_n^*]$, with $\alpha_n=(n+i\gamma\Phi)/\sqrt{2\gamma}$ or, equivalently, $\alpha_n=(n'+i\Phi')/\sqrt{2}$. The two matrices are related by the same transformation used in Sec.~\ref{sec:NonclassicalityGaussian}, namely $V_n^{\rm (c)}=\mathcal{T}^\dagger V_n^{\rm (r)}\mathcal{T}$. In the formal replacement $\hat{a}_n\to\hat{a}$, Eq.~(\ref{Vc_number}) reduces to Eq.~(\ref{Vc_gaussian}).

The eigenvalues of $V_n^{\rm (r)}$ and $V_n^{\rm (c)}$ coincide. They determine the smallest noise $\Lambda_{n,{\rm sm}}\le 1/2$ and the largest noise $\Lambda_{n,{\rm lg}}\ge 1/2$ accessible by local rotations in the scaled $n'$--$\Phi'$ plane. As before, we define $\lambda_{n,{\rm sm}}=2\Lambda_{n,{\rm sm}}$ and $\lambda_{n,{\rm lg}}=2\Lambda_{n,{\rm lg}}$, so that comparison with unity directly reveals squeezing relative to the SQL.

In a general number-phase-squeezed state, the minimum noise need not lie along either $\hat{n}'$ or $\hat{\Phi}'$; instead it may lie along a rotated operator $\hat{a}_{n,\phi}=e^{i\phi}\hat{a}_n$, entirely analogous to $\hat{x}_\phi$ for quadrature squeezing. We therefore ask whether the eigenvalues of $V_n^{\rm (c)}$ or $V_n^{\rm (r)}$ remain unchanged when the state is rotated as $|\psi_\phi\rangle=\exp(i\hat{a}_n^\dagger\hat{a}_n\phi)|\psi\rangle$. The answer is yes: just as in the Gaussian case, $\lambda_{n,{\rm sm}}$ and $\lambda_{n,{\rm lg}}$ are invariant under such intra-mode rotations. This means that the covariance-matrix eigenvalues automatically pick out the optimal local directions, without any manual search.

The rotation $\hat{R}_n(\phi)=\exp(i\hat{a}_n^\dagger\hat{a}_n\phi)$ is physically distinct from the ordinary $x$--$p$ phase-space rotation generated by $\hat{a}_\phi=\hat{a}e^{i\phi}$. The latter leaves the noises $\langle(\Delta\hat{n}')^2\rangle$ and $\langle(\Delta\hat{\Phi}')^2\rangle$ unchanged, whereas $\hat{R}_n(\phi)$ rotates the state in the $n'$--$\Phi'$ plane itself. Under this rotation, the phase of $\langle\hat{a}_n\rangle$ changes by $\phi$ while $|\langle\hat{a}_n\rangle|$ remains fixed. We also find that the generalized displacement operator $\hat{\mathcal{D}}=\exp(\beta\hat{a}_n^\dagger - \beta^*\hat{a}_n)$ shifts the state in the $n'$--$\Phi'$ plane without changing its noise properties.

It is important to recall that the phase operator $\hat{\Phi}$ is unproblematic only in the regime $\langle\hat{n}\rangle\gg 1$~\cite{KitagawaPRA1986numberphase}. More generally, one may use the uncertainty relation
\begin{equation}
\langle(\Delta\hat{n})^2\rangle \: \langle(\Delta\hat{S})^2\rangle \geq \langle\hat{C}\rangle^2/4
\end{equation}
following from $[\hat{n},\hat{S}]=i\hat{C}$, where~\cite{KitagawaPRA1986numberphase}
\begin{equation}
\hat{S}=\frac{1}{2i}(\hat{e}_- - \hat{e}_+) \quad \text{and} \quad \hat{C}=\frac{1}{2}(\hat{e}_- + \hat{e}_+),
\end{equation}
with $\hat{e}_-=(\hat{n}+1)^{-1/2}\hat{a}$ and $\hat{e}_+=\hat{e}_-^\dagger$. In that setting we evaluate Eqs.~(\ref{Vr_number}) and~(\ref{Vc_number}) with the scaled operator
\begin{equation}
\hat{a}_n=\frac{1}{\sqrt{2}} \left( \hat{n} + i \gamma \frac{\hat{S}}{\langle\hat{C}\rangle} \right)/\sqrt{\gamma},
\label{anS}
\end{equation}
that is, with the replacement $\hat{\Phi}\to \hat{S}/\langle\hat{C}\rangle$~\cite{PeggJModOpt1990Intelligent}. The invariance properties described above remain valid in this generalized formulation.

\small{\bf Outlook.---} The number-phase formulation naturally raises several open questions. Can one define a class of states analogous to Gaussian states through a Gaussian-type characteristic function,
\begin{equation}
\chi(n_{1,2}',\Phi_{1,2}')=\exp \left( -\frac{1}{2} u_n^\dagger V^{\rm (r)}_{n_1,n_1} u_n \right),
\label{CharacteristicR}
\end{equation}
with $u_n^\dagger=[n_1',\Phi_1',n_2',\Phi_2']$, or equivalently through
\begin{equation}
\chi(n_{1,2}',\Phi_{1,2}')=\exp \left( -\frac{1}{2} y_n^\dagger V^{\rm (c)}_{n_1,n_1} y_n \right),
\label{CharacteristicC}
\end{equation}
where $y^\dagger = [\alpha_{n_1}^*,\alpha_{n_1},\alpha_{n_2}^*,\alpha_{n_2}]$? If so, can one define a correspondingly simple nonclassical depth $\tau$ for such states? And can the remaining SMNc and TME be separated by setting the generalized correlation block to zero, in direct analogy with Ref.~\cite{ge2015conservation} and Eq.~(\ref{tauent})? We leave these questions for future work.

\subsection{Role of the extra term in the Schr\"odinger--Robertson inequality} \label{sec:extraterm_number}

In Sec.~\ref{sec:extratermGaussian} we showed that, for quadrature variables, the extra term in the SR inequality compensates for nonoptimal mixing directions. The same mechanism also appears for number-phase-squeezed-like states. In this setting the HZ criterion~\cite{Hillery&ZubairyPRL2006} is known to be particularly effective for superpositions of number~(Fock) states~\cite{NhaPRA2006Fock_states}, and Ref.~\cite{Nha&ZubairyPRL2008} derived a stronger criterion by using the SR inequality:
\begin{equation}
(\langle (\Delta\hat{u})^2\rangle + 1)\: (\langle (\Delta\hat{v})^2\rangle + 1) <  \langle\hat{n}_1+\hat{n}_2\rangle + 
\langle\Delta\hat{u}\: \Delta\hat{v}\rangle_{\scaleto{\rm S}{3 pt}}^2
\label{SRineqN}
\end{equation}
with
\begin{eqnarray}
\tilde{u}=\hat{a}_2^\dagger \hat{a}_1 + \hat{a}_1^\dagger\hat{a}_2,
\label{u_n_NhaZub}
\\
\tilde{v}=i(\hat{a}_2^\dagger \hat{a}_1 - \hat{a}_1^\dagger \hat{a}_2 ).
\label{v_n_NhaZub}
\end{eqnarray}
Here the term $\langle\Delta\hat{u}\: \Delta\hat{v}\rangle_{\scaleto{\rm S}{3 pt}}^2$ is precisely the extra SR contribution. Criterion~(\ref{SRineqN}) follows from the PT of the observables $\hat{u}=\hat{a}_2^\dagger \hat{a}_1^\dagger + \hat{a}_1\hat{a}_2$ and $\hat{v}=i(\hat{a}_2^\dagger \hat{a}_1^\dagger - \hat{a}_1\hat{a}_2)$, and it is stronger than both the HZ criterion and the condition in Ref.~\cite{NhaPRA2006Fock_states}.

Consider now an eigenstate of $\hat{E}_n$ from Eq.~(\ref{Enoper}), or of its $\hat{S}/\langle C\rangle$ version. Such a state exhibits number squeezing along $\hat{n}$ for $r=\gamma'/\gamma<1$ and phase squeezing along $\hat{\Phi}$ for $r>1$. Accordingly, the smallest eigenvalue of the covariance matrix~(\ref{Vc_number}) equals $Q_n=\langle(\Delta\hat{n})^2\rangle/\langle\hat{n}\rangle$ for $r<1$ and equals $Q_{\Phi}=4\langle\hat{n}\rangle\langle(\Delta\hat{\Phi})^2\rangle$ for $r>1$. When such an unrotated state is mixed with a coherent state at a BS, the extra term vanishes, $\langle\Delta\tilde{u}\Delta\tilde{v}\rangle_{\scaleto{\rm S}{3 pt}}=0$, so the HUR and SR versions are equally strong.

The situation changes once the eigenstate is rotated in the $n'$--$\Phi'$ plane. As discussed in Sec.~\ref{sec:SMNcnumber}, the true minimum noise $\lambda_{n,{\rm sm}}$ is invariant under this rotation, but the quantities $Q_n$ and $Q_{\Phi}$ generally increase because the squeezing no longer lies along the bare $\hat{n}$ or $\hat{\Phi}$ axes. In that case the collective variables in Eqs.~(\ref{u_n_NhaZub}) and~(\ref{v_n_NhaZub}) are no longer optimal, exactly as in the quadrature case when one mixes $\hat{x}_{1,2}$ instead of the optimum quadratures $\hat{x}_{\phi_{1,2}}$.

\begin{figure}
\includegraphics[width=0.5\textwidth]{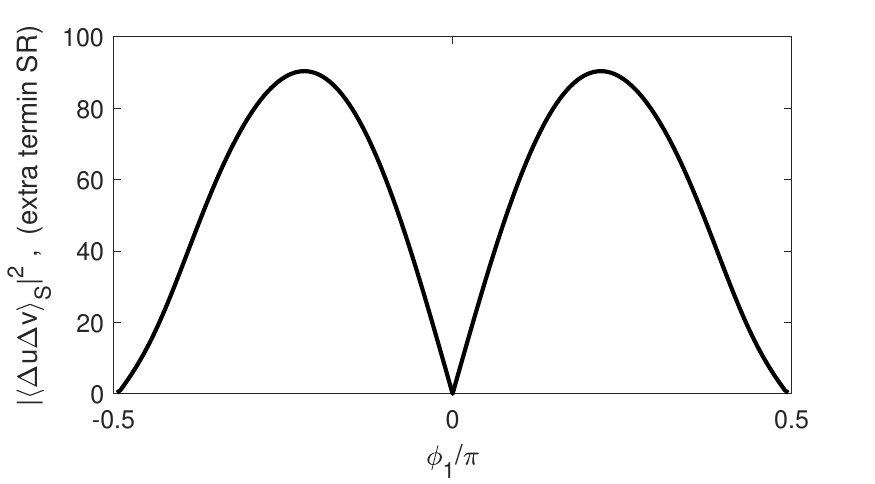}
\caption{Role of the extra term in the SR-based entanglement criterion for number-phase-squeezed-like states. The second mode is already aligned with its minimum-noise direction in the $n_2$--$\Phi_2$ plane. When the first mode is also aligned, $\phi_1=0$, the extra term vanishes. As $\hat{n}_1$ is rotated away from the minimum-noise direction, the extra term grows and compensates for the inefficient use of local nonclassicality in mode~1.}
\label{fig:extra_termN}
\end{figure}

The extra term in Eq.~(\ref{SRineqN}) then compensates for this inefficiency. To illustrate the point, we mix an eigenstate of $\hat{E}_n$ with $r=5/7$ with a coherent state at a BS, then rotate that eigenstate as $|\psi_r(\phi)\rangle=\hat{R}_n(\phi)|\psi_r\rangle$. Figure~\ref{fig:extra_termN} shows the resulting value of $\langle\Delta\tilde{u}\Delta\tilde{v}\rangle_{\scaleto{\rm S}{3 pt}}$ as a function of the rotation angle. For $\phi=0$, the minimum noise lies along $\hat{n}$ and the extra term vanishes. It increases up to $\phi=\pi/4$, where the minimum-noise direction lies halfway between $\hat{n}'$ and $\hat{\Phi}'$, and decreases again as the optimal direction approaches $\hat{\Phi}'$, vanishing once more at $\phi=\pi/2$.

This provides the direct analogue of the Gaussian case: the SR covariance term measures the squeezing missed by a nonoptimal choice of collective variables in the $n$--$\Phi$ plane.

\subsection{Simon-like criterion for number-phase-squeezed states} \label{sec:Simon_number}

In Sec.~\ref{sec:extratermGaussian} we emphasized that Simon's criterion~\cite{SimonPRL2000} automatically optimizes over the local quadrature directions because it is built from quantities that are invariant under local canonical transformations. Since the operators $\hat{a}_n$ introduced above behave algebraically like the standard annihilation operator, it is natural to ask whether Simon's construction can be transplanted to the $n$--$\Phi$ setting. The answer is yes.

Define
\begin{eqnarray}
\mathcal{Q}_1=(\hat{a}_{n,1}^\dagger +\hat{a}_{n,1})/\sqrt{2} , \quad
\mathcal{P}_1=i(\hat{a}_{n,1}^\dagger -\hat{a}_{n,1})/\sqrt{2}, \qquad
\label{Q1P1}
\\
\mathcal{Q}_2=(\hat{a}_{n,2}^\dagger +\hat{a}_{n,2})/\sqrt{2} , \quad
\mathcal{P}_2=i(\hat{a}_{n,2}^\dagger -\hat{a}_{n,2})/\sqrt{2}, \qquad
\label{Q2P2}
\end{eqnarray}
with $\hat{a}\to\hat{a}_n$, where $\hat{a}_n$ is given by Eq.~(\ref{an}) or Eq.~(\ref{anS}). Following the logic of Ref.~\cite{SimonPRL2000}, and defining $\hat{\xi}=[\hat{\mathcal{Q}}_1,\hat{\mathcal{P}}_1,\hat{\mathcal{Q}}_2,\hat{\mathcal{P}}_2]$, one introduces the real covariance matrix
\begin{eqnarray}
V_n^{\rm (r)}=\begin{bmatrix}
    A &  C &  \\
     C^T & B
\end{bmatrix}
\label{Vr_Simon}
\end{eqnarray}
with $V^{\rm (r)}_{n,ij}=\langle \{ \Delta\hat{\xi}_i, \Delta\hat{\xi}_j \} \rangle/2$. For separable states this matrix must satisfy
\begin{eqnarray}
\mu =&& \det{A} \det{B}  + \left(\frac{1}{4} - |\det C| \right)^2 - {\rm tr}(AJCJBJC^TJ) \nonumber \\
-&&\frac{1}{4}(\det A+\det B) \geq 0
\label{Simon_crit_number}
\end{eqnarray}
where $J=[0,1;-1,0]$ and $A$, $B$, and $C$ are $2\times2$ blocks.

Just as in the quadrature case, this Simon-like criterion is invariant under intra-mode rotations and therefore performs the local optimization automatically. In that sense it plays, for number-phase-squeezed-like states, the same role that the SPH criterion plays for quadrature-squeezed Gaussian states. By contrast, the criterion~(\ref{SRineqN}) discussed above depends explicitly on the particular collective variables chosen in Eqs.~(\ref{u_n_NhaZub}) and~(\ref{v_n_NhaZub}).

\begin{figure}
\includegraphics[width=0.5\textwidth]{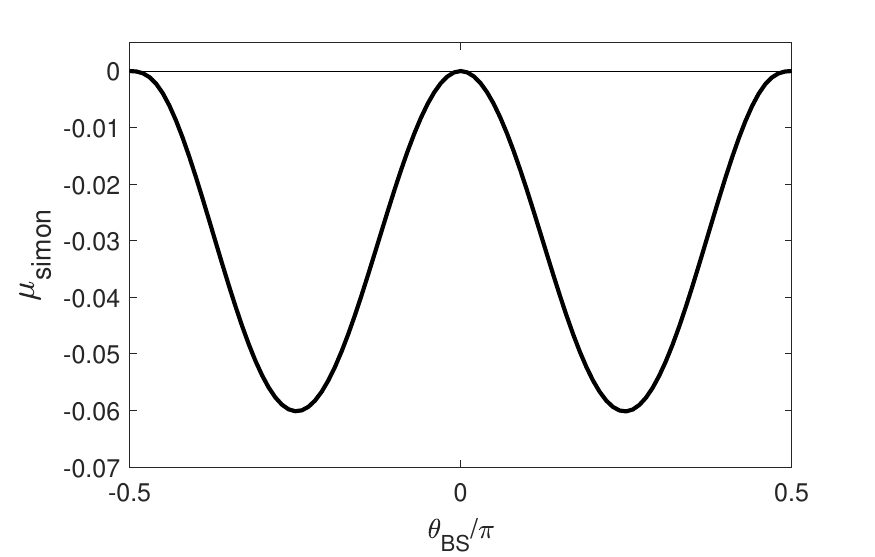}
\caption{Simon-like criterion for number-phase-squeezed-like states. A number-squeezed state is mixed with a coherent state at a beam splitter. The quantity $\mu$ in Eq.~(\ref{Simon_crit_number}) automatically optimizes over intra-mode rotations in the $n_1'$--$\Phi_1'$ and $n_2'$--$\Phi_2'$ planes.}
\label{fig:SimonCritN}
\end{figure}

Under partial transposition one has $\hat{\mathcal{P}}_2 \to - \hat{\mathcal{P}}_2$, exactly as $\hat{p}_2\to-\hat{p}_2$ in the quadrature setting. Figure~\ref{fig:SimonCritN} shows $\mu$ from Eq.~(\ref{Simon_crit_number}) for a number-squeezed state mixed with a coherent state at a BS. The criterion clearly captures the entanglement while automatically optimizing over rotations in the $n_1'$--$\Phi_1'$ and $n_2'$--$\Phi_2'$ planes.

If one adopts a Gaussian-state viewpoint built from $\hat{a}_n$ rather than $\hat{a}$, the logarithmic negativity~\cite{adesso2004extremal,Vidal:02}, which is an entanglement monotone~\cite{plenio2005logarithmic}, can likewise be extended to number-phase-squeezed-like states.

Finally, the same strategy can be generalized to other classes of states by choosing the operators that naturally encode their nonclassicality. For example, for amplitude-squeezed-like states one may define
\begin{equation}
\hat{A}_{\rm amp} = \hat{Y}_1 + i\gamma\hat{Y}_2,
\end{equation}
with $\hat{Y}_1=\hat{a}^\dagger{}^2+\hat{a}^2$ and $\hat{Y}_2=i(\hat{a}^\dagger{}^2- \hat{a}^2)$. In all such cases, a Simon-like construction has the advantage of performing the local optimization automatically.

\subsection{Schr\"odinger--Robertson criterion for $\hat{n}$-$\hat{\Phi}$ variables} \label{sec:SRNPhi}

For completeness, we also present a TME criterion written directly in terms of $\hat{n}'$ and $\hat{\Phi}'$. The construction parallels the DGCZ approach for quadratures: we simply replace $\hat{x}_{1,2}$ by $\hat{n}_{1,2}$ and $\hat{p}_{1,2}$ by $\gamma_{1,2}\hat{\Phi}_{1,2}$.

Choosing $\hat{u}=\hat{a}_2^\dagger \hat{a}_1^\dagger + \hat{a}_1\hat{a}_2$ and $\hat{v}=i(\hat{a}_2^\dagger \hat{a}_1^\dagger - \hat{a}_1\hat{a}_2)$ in the SR inequality~(\ref{SR_ineq}) leads, under PT, to the separability condition~(\ref{SRineqN}). Alternatively, one may use variables analogous to Eqs.~(\ref{uDGCZ}) and~(\ref{vDGCZ}),
\begin{eqnarray}
\hat{u}=\cos\theta \, \hat{n}_1 + \sin\theta \, \hat{n}_2,
\\
\hat{v}=\cos\theta\,\gamma_1\hat{\Phi}_1 + \sin\theta\,\gamma_2\hat{\Phi}_2
\label{vPhi1Phi2}
\end{eqnarray}
and obtain
\begin{eqnarray}
\langle(\Delta\hat{u})^2\rangle \langle(\Delta\tilde{v})^2\rangle \geq && (\cos^2\theta \, \langle\hat{n}_1\rangle+\sin^2\theta \, \langle\hat{n}_2\rangle)^2
\nonumber \\
&&+\langle\Delta\hat{u}\Delta\tilde{v}\rangle_{\scaleto{\rm S}{3 pt}}^2,
\label{SRNPhi}
\end{eqnarray}
where $\tilde{v}=\cos\theta\,\gamma_1\hat{\Phi}_1 - \sin\theta\,\gamma_2\hat{\Phi}_2$ is the PT-transformed version of Eq.~(\ref{vPhi1Phi2}). For comparison, the stronger product form of the criterion by Raymer \textit{et al.}~\cite{RaymerPRA2003} yields
\begin{equation}
\langle(\Delta\hat{u})^2\rangle \langle(\Delta\tilde{v})^2\rangle \geq 4\cos^2\theta\sin^2\theta \langle\hat{n}_1\rangle \langle\hat{n}_2\rangle,
\label{uvprod_separable}
\end{equation}
which is weaker than the first term on the right-hand side of Eq.~(\ref{SRNPhi}), since $(a-b)^2=a^2+b^2-2ab>0$.

\subsection{Noise area in the $\hat{n}'$-$\hat{\Phi}'$ plane and entanglement} \label{sec:noisearea_nPhi}

In Sec.~\ref{sec:ObservNA} we argued that the DGCZ criterion~\cite{DGCZ_PRL2000,Mancini&TombesiPRL2002_DGCZ_product} can be interpreted as a search for a back-rotated noise area below unity in the quadrature plane. Here we present the analogous observation for number-phase-squeezed-like states: the Hillery--Zubairy~(HZ) criterion~\cite{Hillery&ZubairyPRL2006} can be viewed as a weaker, sum-type version of a search for
$\Omega_n=\frac{\langle(\Delta \hat{n}_1)^2\rangle}{\langle\hat{n}_1\rangle} \: \frac{\langle(\Delta \hat{n}_2)^2\rangle}{\langle\hat{n}_2\rangle} <1$.

The HZ criterion,
$\langle \hat{a}_1^\dagger \hat{a}_1 \hat{a}_2^\dagger \hat{a}_2 \rangle < |\langle \hat{a}_2^\dagger \hat{a}_1 \rangle|^2 \Rightarrow$ inseparable,
can be derived by analyzing the sum of spin variances $\langle(\Delta\hat{S}_x)^2\rangle + \langle (\Delta\hat{S}_y)^2 \rangle$~\footnote{The HZ criterion is a subset of the more general moment conditions of Shchukin and Vogel~\cite{ShchukinVogelPRL2005}.}. Here the pseudo-spin operators are $\hat{S}_x=(\hat{S}_++\hat{S}_-)/2$, $\hat{S}_y=-i(\hat{S}_+-\hat{S}_-)/2$, with $\hat{S}_+=\hat{a}_2^\dagger \hat{a}_1$ and $\hat{S}_-=\hat{S}_+^\dagger$, while $\hat{N}_+=\hat{a}_1^\dagger \hat{a}_1+\hat{a}_2^\dagger \hat{a}_2$ is the total photon number.

For separable states, $\hat{\rho}_{1,2}=\sum_k P_k \,\rho_1^{ (k)}\otimes \rho_2^{ (k)}$, one has
\begin{equation}
\mu_{\scaleto{\rm HZ}{3 pt}} \to \langle(\Delta\hat{S}_x)^2\rangle + \langle (\Delta\hat{S}_y)^2 \rangle  \geq
\langle \hat{N}_+ \rangle/2 \: ,
\label{HZSxSy}
\end{equation}
whose violation witnesses TME. This criterion is especially effective for superpositions of number-squeezed-like states~\cite{NhaPRA2006Fock_states}. A product-form version can also be obtained from PT of a HUR-type inequality~\cite{nha2007entanglement}.

To expose the noise area structure, we rotate Eq.~(\ref{HZSxSy}) back by the optimal BS angle $\theta_{\scaleto{\rm BS}{3 pt}}=-\pi/4$. Under this rotation, $\hat{S}_x \to \hat{S}_{\bf z}$ while $\hat{S}_y$ and $\hat{N}_+$ remain invariant, so the rotated HZ criterion becomes
\begin{eqnarray}
\mu_{\scaleto{\rm HZ}{3 pt}}(\theta_{\scaleto{\rm BS}{3 pt}}=-\pi/4)  \to  \langle(\Delta\hat{S}_{\bf z})^2\rangle + \langle (\Delta\hat{S}_y)^2 \rangle  \geq
\langle \hat{N}_+ \rangle/2 \: . \qquad
\label{AreaSzSy}
\end{eqnarray}
We now show that this rotated form is a nonclassicality condition that mirrors the weaker, sum-type version of the noise area criterion in the $n$ plane.

Indeed, Eq.~(\ref{AreaSzSy}) can be rewritten as
\begin{eqnarray}
&&\mu_{\scaleto{\rm HZ}{3 pt}}(\theta_{\scaleto{\rm BS}{3 pt}}=-\pi/4) \to \langle(\Delta\hat{S}_z)^2\rangle \geq \nonumber \\
 &&\langle\hat{N}_+\rangle/4  -\int d^2\alpha_1d^2\alpha_2\: P(\alpha_1,\alpha_2) \left[ \frac{i}{2}(\alpha_2^*\alpha_1-\alpha_1^*\alpha_2) - \langle\hat{S}_y\rangle \right]^2
 \nonumber \\
\label{AreaSzSy2}
\end{eqnarray}
using the normal-ordered form
\begin{eqnarray}
&&\langle(\Delta\hat{S}_y)^2\rangle = \langle\hat{N}_+\rangle/4 \nonumber \\
&&+\int d^2\alpha_1d^2\alpha_2\: P(\alpha_1,\alpha_2) \left[ \frac{i}{2}(\alpha_2^*\alpha_1-\alpha_1^*\alpha_2) - \langle\hat{S}_y\rangle \right]^2.
\qquad
\end{eqnarray}
Therefore, violation of the rotated HZ criterion implies the weaker noise condition
\begin{equation}
\mu_{\scaleto{\rm HZ}{3 pt}}(\theta_{\scaleto{\rm BS}{3 pt}}=-\pi/4)  \xrightarrow[\text{form of}]{\text{weaker}}
 \langle(\Delta\hat{n}_1)^2\rangle + \langle(\Delta\hat{n}_2)^2\rangle < \langle\hat{n}_1\rangle + \langle\hat{n}_2 \rangle.
\label{noise_area_num_sum}
\end{equation}
This is the direct analogue of the relation between the DGCZ sum criterion and the stronger product criterion of Mancini \textit{et al.}~\cite{Mancini&TombesiPRL2002_DGCZ_product}. In the present case, Eq.~(\ref{noise_area_num_sum}) is the sum-type subset of
\begin{equation}
\Omega_n=\frac{\langle(\Delta \hat{n}_1)^2\rangle}{\langle\hat{n}_1\rangle} \: \frac{\langle(\Delta \hat{n}_2)^2\rangle}{\langle\hat{n}_2\rangle} <1
\label{Omegan}
\end{equation}
which is the genuine product-form noise area condition.

One may push the analogy a step further and ask whether the $\theta_{\scaleto{\rm BS}{3 pt}}=\pi/4$-rotated version of Eq.~(\ref{Omegan}) itself yields a new TME criterion. This would amount to examining
\begin{equation}
\langle[\Delta(\hat{N}_++\hat{S}_x)]^2\rangle\: \langle[\Delta(\hat{N}_+ -\hat{S}_x)]^2\rangle  {\stackrel{?}{\geq}}
\langle\hat{N}_+\rangle^2 -\langle\hat{S}_x\rangle^2.
\label{TMEnoise_area_num}
\end{equation}
We have not succeeded in proving this analytically by inserting $\hat{\rho}_{12}=\sum_k P_k \hat{\rho}_1^{(k)} \hat{\rho}_2^{(k)}$ and applying Cauchy--Schwarz inequalities, although our numerical tests strongly suggest that Eq.~(\ref{TMEnoise_area_num}) or a close variant does behave as a TME criterion. In practice, such a criterion would likely be superseded by the Simon-like condition introduced in Sec.~\ref{sec:Simon_number}, which is expected to be stronger. Even so, understanding whether Eq.~(\ref{TMEnoise_area_num}) can be proved rigorously would further clarify the structural relation between noise area conditions and entanglement witnesses.

\section{Summary and discussion} \label{sec:summary}

Nonclassicality conditions derived from the positivity of the Glauber--Sudarshan $P(\alpha_1,\alpha_2)$ function~\cite{ScullyZubairyBook} are powerful but intrinsically ambiguous as entanglement witnesses, because the observed nonclassicality may originate from TME, from local SMNc, or from both. The main purpose of this paper has been to clarify how widely used TME criteria navigate that ambiguity.

The central organizing idea is the \emph{noise area}: the product of the minimum local noises of the two modes, for example $\Omega=\langle(\Delta \hat{x}_1)^2\rangle \langle(\Delta \hat{x}_2)^2\rangle$ or $\Omega_n=\frac{\langle(\Delta \hat{n}_1)^2\rangle}{\langle \hat{n}_1 \rangle} \frac{\langle(\Delta \hat{n}_2)^2\rangle}{\langle \hat{n}_2 \rangle}$. For Gaussian states, the beam splitter formulas of Refs.~\cite{Tahira:09,Li:06} show that the entanglement generated at the output is fixed by the input noise area. We then showed geometrically, through Fig.~\ref{fig:geometry}, and analytically, through the quantity $\tau_{\rm ent}$, how the increase in output noise area reflects the conversion of local nonclassicality into entanglement.

A second main result concerns PT-based criteria derived from the SR inequality. We showed, for both quadrature variables and number-phase-squeezed-like states, that the extra covariance term compensates for nonoptimal collective variables: whenever the criterion mixes directions that are not aligned with the local minimum-noise axes, the missing local squeezing reappears through the extra SR term. Simon's criterion is stronger precisely because it avoids this problem by performing the local optimization automatically. This observation motivated our Simon-like criterion for number-phase-squeezed-like states.

The third main result is the structural link between optimized entanglement criteria and the noise area. For Gaussian states, numerical calculations indicate that the optimized product-form DGCZ quantity,
$\Omega_{\scaleto{\rm DGCZ}{4 pt}}^{\rm (min)}=\left[(\Delta\tilde{u})^2(\Delta\tilde{v})^2\right]^{\rm min}_{\phi_1,\phi_2}$,
coincides with the beam splitter input noise area that would generate the same logarithmic negativity as the given entangled state. In that sense, optimized DGCZ-type criteria behave as searches for a back-rotated noise area below unity. We further argued that the transformation $p_2\to -p_2$ preserves this nonclassicality interpretation because it preserves normal ordering, even though the same algebraic transformation is also used in PT-based separability tests for a different physical reason.

For number-phase-squeezed-like states we found an analogous picture. The covariance-matrix eigenvalues in the $n$--$\Phi$ plane are invariant under local rotations and displacements, the SR extra term again compensates for nonoptimal directions, and the rotated HZ criterion emerges as a weaker, sum-type version of the product noise area condition $\Omega_n<1$. This strongly suggests that the noise area viewpoint is not restricted to Gaussian quadrature variables.

Several questions remain open. Can one formulate a genuinely Gaussian theory in the number-phase setting, with an equally simple notion of nonclassical depth? Can one separate residual SMNc and TME there as cleanly as in the Gaussian beam splitter case? And can the candidate criterion in Eq.~(\ref{TMEnoise_area_num}) be proved analytically? We hope that the structural viewpoint developed here will make such questions easier to approach and, more broadly, will help organize future work on experimentally useful entanglement criteria.

\begin{acknowledgments}
This work is supported by the T\"UB\.ITAK 1001 Grant No.~117F118 and the T\"UBA-GEB\.IP 2017 fund. We gratefully thank M. Suhail Zubairy and Peter Zoller for their hospitality. We also thank Michael G. Raymer and Nathan Killoren for illuminating discussions.
\end{acknowledgments}


\appendix 

\section{Derivation of inequality (\ref{uvprod_separable})} \label{appA}

The derivation follows the method of Mancini \textit{et al.}~\cite{Mancini&TombesiPRL2002_DGCZ_product} and Raymer \textit{et al.}~\cite{RaymerPRA2003}.

Consider the linearly mixed operators
\begin{subequations}
\begin{align}
\hat{u}&=\alpha \hat{A}_1 + \beta \hat{A}_2 \label{uva} \\
\hat{v}&=\alpha \hat{B}_1 \pm \beta \hat{B}_2 \; \label{uvb},
\end{align}
\end{subequations}
where $\hat{A}_1$, $\hat{B}_1$ belong to subsystem~1 and $\hat{A}_2$, $\hat{B}_2$ belong to subsystem~2. The coefficients $\alpha$ and $\beta$ are arbitrary real numbers.

For separable states we insert the decomposition
\begin{equation}
\hat{\rho}=\sum_i P_i \: \hat{\rho}_i^{(1)} \otimes \hat{\rho}_i^{(2)}
\label{density_martix}
\end{equation}
into the moments of $\hat{u}$ and obtain
\begin{equation}
\langle \hat{u}^2\rangle = \alpha^2\sum_i P_i \langle \hat{A}_1^2 \rangle_i + \beta^2\sum_i P_i \langle \hat{A}_2^2 \rangle_i
+ 2\alpha\beta \sum_i P_i \langle \hat{A}_1\rangle_i \langle \hat{A}_2\rangle_i
\label{eq:u2}
\end{equation}
while
\begin{equation}
\sum_i P_i \langle \hat{u}\rangle_i^2 =
\sum_i P_i \left(  \alpha^2\langle \hat{A}_1\rangle_i^2 + \beta^2\langle \hat{A}_2\rangle_i^2 + 2\alpha\beta\langle \hat{A}_1\rangle_i \langle \hat{A}_2\rangle_i \right)
\label{eq:u_mean2}
\end{equation}
Here $\langle\,\rangle_i$ denotes the quantum expectation value in the state that occurs with classical probability $P_i$. The corresponding expressions for $\hat{v}$ are obtained analogously.

Subtracting Eq.~(\ref{eq:u_mean2}) from Eq.~(\ref{eq:u2}) yields
\begin{eqnarray}
\langle(\Delta \hat{u})^2\rangle =&& \alpha^2\sum_i P_i \langle(\Delta\hat{A}_1)^2\rangle_i + \beta^2\sum_i P_i \langle(\Delta\hat{A}_2)^2\rangle_i
\nonumber \\
&&+\sum_i P_i \langle\hat{u}\rangle_i^2 \: - \: \langle \hat{u}\rangle^2 \: .
\label{Delta_u_Pi}
\end{eqnarray}
The second line is nonnegative by the Cauchy--Schwarz inequality,
\begin{equation}
\Big(\sum_i P_i \Big) \Big(\sum_i P_i \langle \hat{u}\rangle_i^2 \Big) \geq
\Big(\sum_i P_i |\langle\hat{u}\rangle_i|\Big)^2 \: .
\end{equation}
Therefore,
\begin{equation}
\langle(\Delta \hat{u})^2\rangle \: \geq \: \alpha^2\sum_i P_i \langle(\Delta\hat{A}_1)^2\rangle_i + \beta^2\sum_i P_i \langle(\Delta\hat{A}_2)^2\rangle_i
\: .
\label{Delta_u_gr}
\end{equation}
An identical inequality holds for $\langle(\Delta \hat{v})^2\rangle$ with $\hat{A}_{1,2}$ replaced by $\hat{B}_{1,2}$.

Using the elementary identity $a^2+b^2\geq 2|a||b|$ on the right-hand side of Eq.~(\ref{Delta_u_gr}), one obtains
\begin{eqnarray}
\langle(\Delta \hat{u})^2\rangle \langle(\Delta \hat{v})^2\rangle \geq
&&4\sum_i P_i |\alpha\beta| \sqrt{\langle(\Delta\hat{A}_1)^2\rangle_i \langle(\Delta\hat{A}_2)^2\rangle_i} \nonumber \\
\times &&\sum_i P_i |\alpha\beta| \sqrt{\langle(\Delta\hat{B}_1)^2\rangle_i \langle(\Delta\hat{B}_2)^2\rangle_i} \nonumber \\
\quad
\end{eqnarray}
which can be rewritten, again by Cauchy--Schwarz, as
\begin{eqnarray}
\langle(\Delta \hat{u})^2\rangle \langle(\Delta \hat{v})^2\rangle \geq \nonumber \\
\resizebox{1.02\hsize}{!}{$4\alpha^2\beta^2 \left[   \sum_i P_i \left(   \langle(\Delta\hat{A}_1)^2\rangle_i \langle(\Delta\hat{A}_2)^2\rangle_i  \langle(\Delta\hat{B}_1)^2\rangle_i \langle(\Delta\hat{B}_2)^2\rangle_i    \right)     \right]^2$ } \nonumber \\
\quad
\end{eqnarray}
Finally, applying the uncertainty relation within each subsystem gives
\begin{equation}
\langle(\Delta \hat{u})^2\rangle \langle(\Delta \hat{v})^2\rangle \geq \alpha^2\beta^2 C_1 C_2
\label{app_Raymor_prod}
\end{equation}
where $C_i=|\langle[\hat{A_i},\hat{B_i}]\rangle|$. This inequality is stronger than the corresponding sum-form result of Ref.~\cite{RaymerPRA2003}.


\bibliography{bibliography}

\begin{thebibliography}{57}%
\makeatletter
\providecommand \@ifxundefined [1]{%
 \@ifx{#1\undefined}
}%
\providecommand \@ifnum [1]{%
 \ifnum #1\expandafter \@firstoftwo
 \else \expandafter \@secondoftwo
 \fi
}%
\providecommand \@ifx [1]{%
 \ifx #1\expandafter \@firstoftwo
 \else \expandafter \@secondoftwo
 \fi
}%
\providecommand \natexlab [1]{#1}%
\providecommand \enquote  [1]{``#1''}%
\providecommand \bibnamefont  [1]{#1}%
\providecommand \bibfnamefont [1]{#1}%
\providecommand \citenamefont [1]{#1}%
\providecommand \href@noop [0]{\@secondoftwo}%
\providecommand \href [0]{\begingroup \@sanitize@url \@href}%
\providecommand \@href[1]{\@@startlink{#1}\@@href}%
\providecommand \@@href[1]{\endgroup#1\@@endlink}%
\providecommand \@sanitize@url [0]{\catcode `\\12\catcode `\$12\catcode
  `\&12\catcode `\#12\catcode `\^12\catcode `\_12\catcode `\%12\relax}%
\providecommand \@@startlink[1]{}%
\providecommand \@@endlink[0]{}%
\providecommand \url  [0]{\begingroup\@sanitize@url \@url }%
\providecommand \@url [1]{\endgroup\@href {#1}{\urlprefix }}%
\providecommand \urlprefix  [0]{URL }%
\providecommand \Eprint [0]{\href }%
\providecommand \doibase [0]{https://doi.org/}%
\providecommand \selectlanguage [0]{\@gobble}%
\providecommand \bibinfo  [0]{\@secondoftwo}%
\providecommand \bibfield  [0]{\@secondoftwo}%
\providecommand \translation [1]{[#1]}%
\providecommand \BibitemOpen [0]{}%
\providecommand \bibitemStop [0]{}%
\providecommand \bibitemNoStop [0]{.\EOS\space}%
\providecommand \EOS [0]{\spacefactor3000\relax}%
\providecommand \BibitemShut  [1]{\csname bibitem#1\endcsname}%
\let\auto@bib@innerbib\@empty
\bibitem [{\citenamefont {Bernstein}\ and\ \citenamefont
  {Lange}(2017)}]{BernsteinNatureCryptography2017}%
  \BibitemOpen
  \bibfield  {author} {\bibinfo {author} {\bibfnamefont {D.~J.}\ \bibnamefont
  {Bernstein}}\ and\ \bibinfo {author} {\bibfnamefont {T.}~\bibnamefont
  {Lange}},\ }\bibfield  {title} {\bibinfo {title} {Post-quantum
  cryptography},\ }\href {https://doi.org/10.1038/nature23461} {\bibfield
  {journal} {\bibinfo  {journal} {Nature}\ }\textbf {\bibinfo {volume} {549}},\
  \bibinfo {pages} {188 EP } (\bibinfo {year} {2017})}\BibitemShut {NoStop}%
\bibitem [{\citenamefont {Pirandola}\ \emph {et~al.}(2015)\citenamefont
  {Pirandola}, \citenamefont {Eisert}, \citenamefont {Weedbrook}, \citenamefont
  {Furusawa},\ and\ \citenamefont
  {Braunstein}}]{BraunsteinNatureTeleportation2015}%
  \BibitemOpen
  \bibfield  {author} {\bibinfo {author} {\bibfnamefont {S.}~\bibnamefont
  {Pirandola}}, \bibinfo {author} {\bibfnamefont {J.}~\bibnamefont {Eisert}},
  \bibinfo {author} {\bibfnamefont {C.}~\bibnamefont {Weedbrook}}, \bibinfo
  {author} {\bibfnamefont {A.}~\bibnamefont {Furusawa}},\ and\ \bibinfo
  {author} {\bibfnamefont {S.~L.}\ \bibnamefont {Braunstein}},\ }\bibfield
  {title} {\bibinfo {title} {Advances in quantum teleportation},\ }\href
  {https://doi.org/10.1038/nphoton.2015.154} {\bibfield  {journal} {\bibinfo
  {journal} {Nature Photonics}\ }\textbf {\bibinfo {volume} {9}},\ \bibinfo
  {pages} {641 EP } (\bibinfo {year} {2015})},\ \bibinfo {note} {review
  Article}\BibitemShut {NoStop}%
\bibitem [{\citenamefont {Giovannetti}\ \emph {et~al.}(2004)\citenamefont
  {Giovannetti}, \citenamefont {Lloyd},\ and\ \citenamefont
  {Maccone}}]{MeasurementSQLSciene2004}%
  \BibitemOpen
  \bibfield  {author} {\bibinfo {author} {\bibfnamefont {V.}~\bibnamefont
  {Giovannetti}}, \bibinfo {author} {\bibfnamefont {S.}~\bibnamefont {Lloyd}},\
  and\ \bibinfo {author} {\bibfnamefont {L.}~\bibnamefont {Maccone}},\
  }\bibfield  {title} {\bibinfo {title} {Quantum-enhanced measurements: beating
  the standard quantum limit},\ }\href@noop {} {\bibfield  {journal} {\bibinfo
  {journal} {Science}\ }\textbf {\bibinfo {volume} {306}},\ \bibinfo {pages}
  {1330} (\bibinfo {year} {2004})}\BibitemShut {NoStop}%
\bibitem [{\citenamefont {Aasi}\ \emph {et~al.}(2013)\citenamefont {Aasi},
  \citenamefont {Abadie}, \citenamefont {Abbott}, \citenamefont {Abbott},
  \citenamefont {Abbott}, \citenamefont {Abernathy}, \citenamefont {Adams},
  \citenamefont {Adams}, \citenamefont {Addesso}, \citenamefont {Adhikari}
  \emph {et~al.}}]{LIGO2013}%
  \BibitemOpen
  \bibfield  {author} {\bibinfo {author} {\bibfnamefont {J.}~\bibnamefont
  {Aasi}}, \bibinfo {author} {\bibfnamefont {J.}~\bibnamefont {Abadie}},
  \bibinfo {author} {\bibfnamefont {B.}~\bibnamefont {Abbott}}, \bibinfo
  {author} {\bibfnamefont {R.}~\bibnamefont {Abbott}}, \bibinfo {author}
  {\bibfnamefont {T.}~\bibnamefont {Abbott}}, \bibinfo {author} {\bibfnamefont
  {M.}~\bibnamefont {Abernathy}}, \bibinfo {author} {\bibfnamefont
  {C.}~\bibnamefont {Adams}}, \bibinfo {author} {\bibfnamefont
  {T.}~\bibnamefont {Adams}}, \bibinfo {author} {\bibfnamefont
  {P.}~\bibnamefont {Addesso}}, \bibinfo {author} {\bibfnamefont
  {R.}~\bibnamefont {Adhikari}}, \emph {et~al.},\ }\bibfield  {title} {\bibinfo
  {title} {Enhanced sensitivity of the ligo gravitational wave detector by
  using squeezed states of light},\ }\href@noop {} {\bibfield  {journal}
  {\bibinfo  {journal} {Nature Photonics}\ }\textbf {\bibinfo {volume} {7}},\
  \bibinfo {pages} {613} (\bibinfo {year} {2013})}\BibitemShut {NoStop}%
\bibitem [{\citenamefont {Las~Heras}\ \emph {et~al.}(2017)\citenamefont
  {Las~Heras}, \citenamefont {Di~Candia}, \citenamefont {Fedorov},
  \citenamefont {Deppe}, \citenamefont {Sanz},\ and\ \citenamefont
  {Solano}}]{QuantumRadarSciRep2017}%
  \BibitemOpen
  \bibfield  {author} {\bibinfo {author} {\bibfnamefont {U.}~\bibnamefont
  {Las~Heras}}, \bibinfo {author} {\bibfnamefont {R.}~\bibnamefont
  {Di~Candia}}, \bibinfo {author} {\bibfnamefont {K.}~\bibnamefont {Fedorov}},
  \bibinfo {author} {\bibfnamefont {F.}~\bibnamefont {Deppe}}, \bibinfo
  {author} {\bibfnamefont {M.}~\bibnamefont {Sanz}},\ and\ \bibinfo {author}
  {\bibfnamefont {E.}~\bibnamefont {Solano}},\ }\bibfield  {title} {\bibinfo
  {title} {Quantum illumination reveals phase-shift inducing cloaking},\
  }\href@noop {} {\bibfield  {journal} {\bibinfo  {journal} {Scientific
  reports}\ }\textbf {\bibinfo {volume} {7}},\ \bibinfo {pages} {9333}
  (\bibinfo {year} {2017})}\BibitemShut {NoStop}%
\bibitem [{\citenamefont {Bozhevolnyi}\ and\ \citenamefont
  {Mortensen}(2017)}]{bozhevolnyi2017plasmonics}%
  \BibitemOpen
  \bibfield  {author} {\bibinfo {author} {\bibfnamefont {S.~I.}\ \bibnamefont
  {Bozhevolnyi}}\ and\ \bibinfo {author} {\bibfnamefont {N.~A.}\ \bibnamefont
  {Mortensen}},\ }\bibfield  {title} {\bibinfo {title} {Plasmonics for emerging
  quantum technologies},\ }\href@noop {} {\bibfield  {journal} {\bibinfo
  {journal} {Nanophotonics}\ }\textbf {\bibinfo {volume} {6}},\ \bibinfo
  {pages} {1185} (\bibinfo {year} {2017})}\BibitemShut {NoStop}%
\bibitem [{\citenamefont {You}\ \emph {et~al.}(2020)\citenamefont {You},
  \citenamefont {Nellikka}, \citenamefont {De~Leon},\ and\ \citenamefont
  {Maga{\~n}a-Loaiza}}]{you2020multiparticle}%
  \BibitemOpen
  \bibfield  {author} {\bibinfo {author} {\bibfnamefont {C.}~\bibnamefont
  {You}}, \bibinfo {author} {\bibfnamefont {A.~C.}\ \bibnamefont {Nellikka}},
  \bibinfo {author} {\bibfnamefont {I.}~\bibnamefont {De~Leon}},\ and\ \bibinfo
  {author} {\bibfnamefont {O.~S.}\ \bibnamefont {Maga{\~n}a-Loaiza}},\
  }\bibfield  {title} {\bibinfo {title} {Multiparticle quantum plasmonics},\
  }\href@noop {} {\bibfield  {journal} {\bibinfo  {journal} {Nanophotonics}\
  }\textbf {\bibinfo {volume} {1}} (\bibinfo {year} {2020})}\BibitemShut
  {NoStop}%
\bibitem [{\citenamefont {Huck}\ \emph {et~al.}(2009)\citenamefont {Huck},
  \citenamefont {Smolka}, \citenamefont {Lodahl}, \citenamefont {S{\o}rensen},
  \citenamefont {Boltasseva}, \citenamefont {Janousek},\ and\ \citenamefont
  {Andersen}}]{hucPRLk2009squeezedplasmon}%
  \BibitemOpen
  \bibfield  {author} {\bibinfo {author} {\bibfnamefont {A.}~\bibnamefont
  {Huck}}, \bibinfo {author} {\bibfnamefont {S.}~\bibnamefont {Smolka}},
  \bibinfo {author} {\bibfnamefont {P.}~\bibnamefont {Lodahl}}, \bibinfo
  {author} {\bibfnamefont {A.~S.}\ \bibnamefont {S{\o}rensen}}, \bibinfo
  {author} {\bibfnamefont {A.}~\bibnamefont {Boltasseva}}, \bibinfo {author}
  {\bibfnamefont {J.}~\bibnamefont {Janousek}},\ and\ \bibinfo {author}
  {\bibfnamefont {U.~L.}\ \bibnamefont {Andersen}},\ }\bibfield  {title}
  {\bibinfo {title} {Demonstration of quadrature-squeezed surface plasmons in a
  gold waveguide},\ }\href@noop {} {\bibfield  {journal} {\bibinfo  {journal}
  {Physical review letters}\ }\textbf {\bibinfo {volume} {102}},\ \bibinfo
  {pages} {246802} (\bibinfo {year} {2009})}\BibitemShut {NoStop}%
\bibitem [{\citenamefont {Tasgin}\ \emph {et~al.}(2020)\citenamefont {Tasgin},
  \citenamefont {Gunay},\ and\ \citenamefont
  {Zubairy}}]{tasgin2020wavepackets}%
  \BibitemOpen
  \bibfield  {author} {\bibinfo {author} {\bibfnamefont {M.~E.}\ \bibnamefont
  {Tasgin}}, \bibinfo {author} {\bibfnamefont {M.}~\bibnamefont {Gunay}},\ and\
  \bibinfo {author} {\bibfnamefont {M.~S.}\ \bibnamefont {Zubairy}},\
  }\bibfield  {title} {\bibinfo {title} {Nonclassicality and entanglement for
  wave packets},\ }\href@noop {} {\bibfield  {journal} {\bibinfo  {journal}
  {Physical Review A}\ }\textbf {\bibinfo {volume} {101}},\ \bibinfo {pages}
  {062316} (\bibinfo {year} {2020})}\BibitemShut {NoStop}%
\bibitem [{\citenamefont {Tame}\ \emph {et~al.}(2013)\citenamefont {Tame},
  \citenamefont {McEnery}, \citenamefont {{\"O}zdemir}, \citenamefont {Lee},
  \citenamefont {Maier},\ and\ \citenamefont {Kim}}]{tame2013quantum}%
  \BibitemOpen
  \bibfield  {author} {\bibinfo {author} {\bibfnamefont {M.~S.}\ \bibnamefont
  {Tame}}, \bibinfo {author} {\bibfnamefont {K.}~\bibnamefont {McEnery}},
  \bibinfo {author} {\bibfnamefont {{\c{S}}.}~\bibnamefont {{\"O}zdemir}},
  \bibinfo {author} {\bibfnamefont {J.}~\bibnamefont {Lee}}, \bibinfo {author}
  {\bibfnamefont {S.~A.}\ \bibnamefont {Maier}},\ and\ \bibinfo {author}
  {\bibfnamefont {M.}~\bibnamefont {Kim}},\ }\bibfield  {title} {\bibinfo
  {title} {Quantum plasmonics},\ }\href@noop {} {\bibfield  {journal} {\bibinfo
   {journal} {Nature Physics}\ }\textbf {\bibinfo {volume} {9}},\ \bibinfo
  {pages} {329} (\bibinfo {year} {2013})}\BibitemShut {NoStop}%
\bibitem [{\citenamefont {Panahpour}\ \emph {et~al.}(2019)\citenamefont
  {Panahpour}, \citenamefont {Mahmoodpoor},\ and\ \citenamefont
  {Lavrinenko}}]{Lavrinenko2019IndexEnhancement}%
  \BibitemOpen
  \bibfield  {author} {\bibinfo {author} {\bibfnamefont {A.}~\bibnamefont
  {Panahpour}}, \bibinfo {author} {\bibfnamefont {A.}~\bibnamefont
  {Mahmoodpoor}},\ and\ \bibinfo {author} {\bibfnamefont {A.~V.}\ \bibnamefont
  {Lavrinenko}},\ }\bibfield  {title} {\bibinfo {title} {Refraction enhancement
  in plasmonics by coherent control of plasmon resonances},\ }\href@noop {}
  {\bibfield  {journal} {\bibinfo  {journal} {Physical Review B}\ }\textbf
  {\bibinfo {volume} {100}},\ \bibinfo {pages} {075427} (\bibinfo {year}
  {2019})}\BibitemShut {NoStop}%
\bibitem [{\citenamefont {Caglayan}\ \emph {et~al.}(2017)\citenamefont
  {Caglayan}, \citenamefont {Hajian},\ and\ \citenamefont
  {Ozbay}}]{caglayan2017coherence}%
  \BibitemOpen
  \bibfield  {author} {\bibinfo {author} {\bibfnamefont {H.}~\bibnamefont
  {Caglayan}}, \bibinfo {author} {\bibfnamefont {H.}~\bibnamefont {Hajian}},\
  and\ \bibinfo {author} {\bibfnamefont {E.}~\bibnamefont {Ozbay}},\ }\bibfield
   {title} {\bibinfo {title} {Controlling coherence in epsilon-near-zero
  metamaterials (conference presentation)},\ }in\ \href@noop {} {\emph
  {\bibinfo {booktitle} {Metamaterials XI}}},\ Vol.\ \bibinfo {volume} {10227}\
  (\bibinfo {organization} {International Society for Optics and Photonics},\
  \bibinfo {year} {2017})\ p.\ \bibinfo {pages} {102270I}\BibitemShut {NoStop}%
\bibitem [{\citenamefont {Vertchenko}\ \emph {et~al.}(2019)\citenamefont
  {Vertchenko}, \citenamefont {Akopian},\ and\ \citenamefont
  {Lavrinenko}}]{Lavrinenko2019ENZquantumnetworks}%
  \BibitemOpen
  \bibfield  {author} {\bibinfo {author} {\bibfnamefont {L.}~\bibnamefont
  {Vertchenko}}, \bibinfo {author} {\bibfnamefont {N.}~\bibnamefont
  {Akopian}},\ and\ \bibinfo {author} {\bibfnamefont {A.~V.}\ \bibnamefont
  {Lavrinenko}},\ }\bibfield  {title} {\bibinfo {title} {Epsilon-near-zero
  grids for on-chip quantum networks},\ }\href@noop {} {\bibfield  {journal}
  {\bibinfo  {journal} {Scientific reports}\ }\textbf {\bibinfo {volume} {9}},\
  \bibinfo {pages} {1} (\bibinfo {year} {2019})}\BibitemShut {NoStop}%
\bibitem [{\citenamefont {Kim}\ \emph {et~al.}(2002)\citenamefont {Kim},
  \citenamefont {Son}, \citenamefont {Bu\ifmmode~\check{z}\else \v{z}\fi{}ek},\
  and\ \citenamefont {Knight}}]{Kim:02}%
  \BibitemOpen
  \bibfield  {author} {\bibinfo {author} {\bibfnamefont {M.~S.}\ \bibnamefont
  {Kim}}, \bibinfo {author} {\bibfnamefont {W.}~\bibnamefont {Son}}, \bibinfo
  {author} {\bibfnamefont {V.}~\bibnamefont {Bu\ifmmode~\check{z}\else
  \v{z}\fi{}ek}},\ and\ \bibinfo {author} {\bibfnamefont {P.~L.}\ \bibnamefont
  {Knight}},\ }\bibfield  {title} {\bibinfo {title} {Entanglement by a beam
  splitter: Nonclassicality as a prerequisite for entanglement},\ }\href
  {https://doi.org/10.1103/PhysRevA.65.032323} {\bibfield  {journal} {\bibinfo
  {journal} {Phys. Rev. A}\ }\textbf {\bibinfo {volume} {65}},\ \bibinfo
  {pages} {032323} (\bibinfo {year} {2002})}\BibitemShut {NoStop}%
\bibitem [{\citenamefont {Tahira}\ \emph {et~al.}(2009)\citenamefont {Tahira},
  \citenamefont {Ikram}, \citenamefont {Nha},\ and\ \citenamefont
  {Zubairy}}]{Tahira:09}%
  \BibitemOpen
  \bibfield  {author} {\bibinfo {author} {\bibfnamefont {R.}~\bibnamefont
  {Tahira}}, \bibinfo {author} {\bibfnamefont {M.}~\bibnamefont {Ikram}},
  \bibinfo {author} {\bibfnamefont {H.}~\bibnamefont {Nha}},\ and\ \bibinfo
  {author} {\bibfnamefont {M.~S.}\ \bibnamefont {Zubairy}},\ }\bibfield
  {title} {\bibinfo {title} {Entanglement of gaussian states using a beam
  splitter},\ }\href {https://doi.org/10.1103/PhysRevA.79.023816} {\bibfield
  {journal} {\bibinfo  {journal} {Phys. Rev. A}\ }\textbf {\bibinfo {volume}
  {79}},\ \bibinfo {pages} {023816} (\bibinfo {year} {2009})}\BibitemShut
  {NoStop}%
\bibitem [{\citenamefont {Hald}\ \emph {et~al.}(1999)\citenamefont {Hald},
  \citenamefont {S{\o}rensen}, \citenamefont {Schori},\ and\ \citenamefont
  {Polzik}}]{PolzikPRL1999spinsqz}%
  \BibitemOpen
  \bibfield  {author} {\bibinfo {author} {\bibfnamefont {J.}~\bibnamefont
  {Hald}}, \bibinfo {author} {\bibfnamefont {J.}~\bibnamefont {S{\o}rensen}},
  \bibinfo {author} {\bibfnamefont {C.}~\bibnamefont {Schori}},\ and\ \bibinfo
  {author} {\bibfnamefont {E.}~\bibnamefont {Polzik}},\ }\bibfield  {title}
  {\bibinfo {title} {Spin squeezed atoms: a macroscopic entangled ensemble
  created by light},\ }\href@noop {} {\bibfield  {journal} {\bibinfo  {journal}
  {Physical Review Letters}\ }\textbf {\bibinfo {volume} {83}},\ \bibinfo
  {pages} {1319} (\bibinfo {year} {1999})}\BibitemShut {NoStop}%
\bibitem [{\citenamefont {Vitagliano}\ \emph {et~al.}(2018)\citenamefont
  {Vitagliano}, \citenamefont {Colangelo}, \citenamefont {Ciurana},
  \citenamefont {Mitchell}, \citenamefont {Sewell},\ and\ \citenamefont
  {T{\'o}th}}]{TothPRA2018SpinSqz}%
  \BibitemOpen
  \bibfield  {author} {\bibinfo {author} {\bibfnamefont {G.}~\bibnamefont
  {Vitagliano}}, \bibinfo {author} {\bibfnamefont {G.}~\bibnamefont
  {Colangelo}}, \bibinfo {author} {\bibfnamefont {F.~M.}\ \bibnamefont
  {Ciurana}}, \bibinfo {author} {\bibfnamefont {M.~W.}\ \bibnamefont
  {Mitchell}}, \bibinfo {author} {\bibfnamefont {R.~J.}\ \bibnamefont
  {Sewell}},\ and\ \bibinfo {author} {\bibfnamefont {G.}~\bibnamefont
  {T{\'o}th}},\ }\bibfield  {title} {\bibinfo {title} {Entanglement and extreme
  planar spin squeezing},\ }\href@noop {} {\bibfield  {journal} {\bibinfo
  {journal} {Physical Review A}\ }\textbf {\bibinfo {volume} {97}},\ \bibinfo
  {pages} {020301} (\bibinfo {year} {2018})}\BibitemShut {NoStop}%
\bibitem [{\citenamefont {Vidal}(2006)}]{vidal2006concurrence}%
  \BibitemOpen
  \bibfield  {author} {\bibinfo {author} {\bibfnamefont {J.}~\bibnamefont
  {Vidal}},\ }\bibfield  {title} {\bibinfo {title} {Concurrence in collective
  models},\ }\href@noop {} {\bibfield  {journal} {\bibinfo  {journal} {Physical
  Review A}\ }\textbf {\bibinfo {volume} {73}},\ \bibinfo {pages} {062318}
  (\bibinfo {year} {2006})}\BibitemShut {NoStop}%
\bibitem [{\citenamefont {Regula}\ \emph {et~al.}(2018)\citenamefont {Regula},
  \citenamefont {Piani}, \citenamefont {Cianciaruso}, \citenamefont {Bromley},
  \citenamefont {Streltsov},\ and\ \citenamefont
  {Adesso}}]{regula2018converting}%
  \BibitemOpen
  \bibfield  {author} {\bibinfo {author} {\bibfnamefont {B.}~\bibnamefont
  {Regula}}, \bibinfo {author} {\bibfnamefont {M.}~\bibnamefont {Piani}},
  \bibinfo {author} {\bibfnamefont {M.}~\bibnamefont {Cianciaruso}}, \bibinfo
  {author} {\bibfnamefont {T.~R.}\ \bibnamefont {Bromley}}, \bibinfo {author}
  {\bibfnamefont {A.}~\bibnamefont {Streltsov}},\ and\ \bibinfo {author}
  {\bibfnamefont {G.}~\bibnamefont {Adesso}},\ }\bibfield  {title} {\bibinfo
  {title} {Converting multilevel nonclassicality into genuine multipartite
  entanglement},\ }\href@noop {} {\bibfield  {journal} {\bibinfo  {journal}
  {New Journal of Physics}\ } (\bibinfo {year} {2018})}\BibitemShut {NoStop}%
\bibitem [{\citenamefont {Ta{\c{s}}g{\i}n}\ and\ \citenamefont
  {Meystre}(2011)}]{Tasgin&MeystrePRA2011}%
  \BibitemOpen
  \bibfield  {author} {\bibinfo {author} {\bibfnamefont {M.~E.}\ \bibnamefont
  {Ta{\c{s}}g{\i}n}}\ and\ \bibinfo {author} {\bibfnamefont {P.}~\bibnamefont
  {Meystre}},\ }\bibfield  {title} {\bibinfo {title} {Spin squeezing with
  coherent light via entanglement swapping},\ }\href@noop {} {\bibfield
  {journal} {\bibinfo  {journal} {Physical Review A}\ }\textbf {\bibinfo
  {volume} {83}},\ \bibinfo {pages} {053848} (\bibinfo {year}
  {2011})}\BibitemShut {NoStop}%
\bibitem [{\citenamefont {Ge}\ \emph {et~al.}(2015)\citenamefont {Ge},
  \citenamefont {Tasgin},\ and\ \citenamefont {Zubairy}}]{ge2015conservation}%
  \BibitemOpen
  \bibfield  {author} {\bibinfo {author} {\bibfnamefont {W.}~\bibnamefont
  {Ge}}, \bibinfo {author} {\bibfnamefont {M.~E.}\ \bibnamefont {Tasgin}},\
  and\ \bibinfo {author} {\bibfnamefont {M.~S.}\ \bibnamefont {Zubairy}},\
  }\bibfield  {title} {\bibinfo {title} {Conservation relation of
  nonclassicality and entanglement for gaussian states in a beam splitter},\
  }\href@noop {} {\bibfield  {journal} {\bibinfo  {journal} {Physical Review
  A}\ }\textbf {\bibinfo {volume} {92}},\ \bibinfo {pages} {052328} (\bibinfo
  {year} {2015})}\BibitemShut {NoStop}%
\bibitem [{\citenamefont {Arkhipov}\ \emph
  {et~al.}(2016{\natexlab{a}})\citenamefont {Arkhipov}, \citenamefont
  {Pe{\v{r}}ina~Jr}, \citenamefont {Svozil{\'\i}k},\ and\ \citenamefont
  {Miranowicz}}]{arkhipov2016nonclassicality}%
  \BibitemOpen
  \bibfield  {author} {\bibinfo {author} {\bibfnamefont {I.~I.}\ \bibnamefont
  {Arkhipov}}, \bibinfo {author} {\bibfnamefont {J.}~\bibnamefont
  {Pe{\v{r}}ina~Jr}}, \bibinfo {author} {\bibfnamefont {J.}~\bibnamefont
  {Svozil{\'\i}k}},\ and\ \bibinfo {author} {\bibfnamefont {A.}~\bibnamefont
  {Miranowicz}},\ }\bibfield  {title} {\bibinfo {title} {Nonclassicality
  invariant of general two-mode gaussian states},\ }\href@noop {} {\bibfield
  {journal} {\bibinfo  {journal} {Scientific reports}\ }\textbf {\bibinfo
  {volume} {6}},\ \bibinfo {pages} {26523} (\bibinfo {year}
  {2016}{\natexlab{a}})}\BibitemShut {NoStop}%
\bibitem [{\citenamefont {Arkhipov}\ \emph
  {et~al.}(2016{\natexlab{b}})\citenamefont {Arkhipov}, \citenamefont
  {Pe{\v{r}}ina~Jr}, \citenamefont {Pe{\v{r}}ina},\ and\ \citenamefont
  {Miranowicz}}]{arkhipov2016interplay}%
  \BibitemOpen
  \bibfield  {author} {\bibinfo {author} {\bibfnamefont {I.~I.}\ \bibnamefont
  {Arkhipov}}, \bibinfo {author} {\bibfnamefont {J.}~\bibnamefont
  {Pe{\v{r}}ina~Jr}}, \bibinfo {author} {\bibfnamefont {J.}~\bibnamefont
  {Pe{\v{r}}ina}},\ and\ \bibinfo {author} {\bibfnamefont {A.}~\bibnamefont
  {Miranowicz}},\ }\bibfield  {title} {\bibinfo {title} {Interplay of
  nonclassicality and entanglement of two-mode gaussian fields generated in
  optical parametric processes},\ }\href@noop {} {\bibfield  {journal}
  {\bibinfo  {journal} {Physical Review A}\ }\textbf {\bibinfo {volume} {94}},\
  \bibinfo {pages} {013807} (\bibinfo {year} {2016}{\natexlab{b}})}\BibitemShut
  {NoStop}%
\bibitem [{\citenamefont {{\v{C}}ernoch}\ \emph {et~al.}(2018)\citenamefont
  {{\v{C}}ernoch}, \citenamefont {Bartkiewicz}, \citenamefont {Lemr},\ and\
  \citenamefont {Soubusta}}]{vcernoch2018experimental}%
  \BibitemOpen
  \bibfield  {author} {\bibinfo {author} {\bibfnamefont {A.}~\bibnamefont
  {{\v{C}}ernoch}}, \bibinfo {author} {\bibfnamefont {K.}~\bibnamefont
  {Bartkiewicz}}, \bibinfo {author} {\bibfnamefont {K.}~\bibnamefont {Lemr}},\
  and\ \bibinfo {author} {\bibfnamefont {J.}~\bibnamefont {Soubusta}},\
  }\bibfield  {title} {\bibinfo {title} {Experimental tests of coherence and
  entanglement conservation under unitary evolutions},\ }\href@noop {}
  {\bibfield  {journal} {\bibinfo  {journal} {Physical Review A}\ }\textbf
  {\bibinfo {volume} {97}},\ \bibinfo {pages} {042305} (\bibinfo {year}
  {2018})}\BibitemShut {NoStop}%
\bibitem [{\citenamefont {Tasgin}\ and\ \citenamefont
  {Zubairy}(2020)}]{tasgin2020multimode}%
  \BibitemOpen
  \bibfield  {author} {\bibinfo {author} {\bibfnamefont {M.~E.}\ \bibnamefont
  {Tasgin}}\ and\ \bibinfo {author} {\bibfnamefont {M.~S.}\ \bibnamefont
  {Zubairy}},\ }\bibfield  {title} {\bibinfo {title} {Quantifications for
  multimode entanglement},\ }\href@noop {} {\bibfield  {journal} {\bibinfo
  {journal} {Physical Review A}\ }\textbf {\bibinfo {volume} {101}},\ \bibinfo
  {pages} {012324} (\bibinfo {year} {2020})}\BibitemShut {NoStop}%
\bibitem [{\citenamefont {Hillery}\ and\ \citenamefont
  {Zubairy}(2006{\natexlab{a}})}]{hillery2006PRA}%
  \BibitemOpen
  \bibfield  {author} {\bibinfo {author} {\bibfnamefont {M.}~\bibnamefont
  {Hillery}}\ and\ \bibinfo {author} {\bibfnamefont {M.~S.}\ \bibnamefont
  {Zubairy}},\ }\bibfield  {title} {\bibinfo {title} {Entanglement conditions
  for two-mode states: Applications},\ }\href@noop {} {\bibfield  {journal}
  {\bibinfo  {journal} {Physical Review A}\ }\textbf {\bibinfo {volume} {74}},\
  \bibinfo {pages} {032333} (\bibinfo {year} {2006}{\natexlab{a}})}\BibitemShut
  {NoStop}%
\bibitem [{\citenamefont {Miranowicz}\ \emph {et~al.}(2010)\citenamefont
  {Miranowicz}, \citenamefont {Bartkowiak}, \citenamefont {Wang}, \citenamefont
  {Liu},\ and\ \citenamefont {Nori}}]{miranowicz2010testing}%
  \BibitemOpen
  \bibfield  {author} {\bibinfo {author} {\bibfnamefont {A.}~\bibnamefont
  {Miranowicz}}, \bibinfo {author} {\bibfnamefont {M.}~\bibnamefont
  {Bartkowiak}}, \bibinfo {author} {\bibfnamefont {X.}~\bibnamefont {Wang}},
  \bibinfo {author} {\bibfnamefont {Y.-x.}\ \bibnamefont {Liu}},\ and\ \bibinfo
  {author} {\bibfnamefont {F.}~\bibnamefont {Nori}},\ }\bibfield  {title}
  {\bibinfo {title} {Testing nonclassicality in multimode fields: A unified
  derivation of classical inequalities},\ }\href@noop {} {\bibfield  {journal}
  {\bibinfo  {journal} {Physical Review A}\ }\textbf {\bibinfo {volume} {82}},\
  \bibinfo {pages} {013824} (\bibinfo {year} {2010})}\BibitemShut {NoStop}%
\bibitem [{\citenamefont {Gholipour}\ and\ \citenamefont
  {Shahandeh}(2016)}]{gholipour2016entanglement}%
  \BibitemOpen
  \bibfield  {author} {\bibinfo {author} {\bibfnamefont {H.}~\bibnamefont
  {Gholipour}}\ and\ \bibinfo {author} {\bibfnamefont {F.}~\bibnamefont
  {Shahandeh}},\ }\bibfield  {title} {\bibinfo {title} {Entanglement and
  nonclassicality: A mutual impression},\ }\href@noop {} {\bibfield  {journal}
  {\bibinfo  {journal} {Physical Review A}\ }\textbf {\bibinfo {volume} {93}},\
  \bibinfo {pages} {062318} (\bibinfo {year} {2016})}\BibitemShut {NoStop}%
\bibitem [{\citenamefont {Tasgin}(2017)}]{tasgin2017many}%
  \BibitemOpen
  \bibfield  {author} {\bibinfo {author} {\bibfnamefont {M.~E.}\ \bibnamefont
  {Tasgin}},\ }\bibfield  {title} {\bibinfo {title} {Many-particle entanglement
  criterion for superradiantlike states},\ }\href@noop {} {\bibfield  {journal}
  {\bibinfo  {journal} {Physical review letters}\ }\textbf {\bibinfo {volume}
  {119}},\ \bibinfo {pages} {033601} (\bibinfo {year} {2017})}\BibitemShut
  {NoStop}%
\bibitem [{\citenamefont {Tasgin}(2020{\natexlab{a}})}]{tasgin2020single}%
  \BibitemOpen
  \bibfield  {author} {\bibinfo {author} {\bibfnamefont {M.~E.}\ \bibnamefont
  {Tasgin}},\ }\bibfield  {title} {\bibinfo {title} {Single-mode
  nonclassicality criteria via holstein-primakoff transformation},\ }\href@noop
  {} {\bibfield  {journal} {\bibinfo  {journal} {Journal of Physics B: Atomic,
  Molecular and Optical Physics}\ } (\bibinfo {year}
  {2020}{\natexlab{a}})}\BibitemShut {NoStop}%
\bibitem [{\citenamefont {Tasgin}(2020{\natexlab{b}})}]{tasgin2020measuring}%
  \BibitemOpen
  \bibfield  {author} {\bibinfo {author} {\bibfnamefont {M.~E.}\ \bibnamefont
  {Tasgin}},\ }\bibfield  {title} {\bibinfo {title} {Measuring nonclassicality
  of single-mode systems},\ }\href@noop {} {\bibfield  {journal} {\bibinfo
  {journal} {Journal of Physics B: Atomic, Molecular and Optical Physics}\
  }\textbf {\bibinfo {volume} {53}},\ \bibinfo {pages} {175501} (\bibinfo
  {year} {2020}{\natexlab{b}})}\BibitemShut {NoStop}%
\bibitem [{\citenamefont {Duan}\ \emph {et~al.}(2000)\citenamefont {Duan},
  \citenamefont {Giedke}, \citenamefont {Cirac},\ and\ \citenamefont
  {Zoller}}]{DGCZ_PRL2000}%
  \BibitemOpen
  \bibfield  {author} {\bibinfo {author} {\bibfnamefont {L.-M.}\ \bibnamefont
  {Duan}}, \bibinfo {author} {\bibfnamefont {G.}~\bibnamefont {Giedke}},
  \bibinfo {author} {\bibfnamefont {J.}~\bibnamefont {Cirac}},\ and\ \bibinfo
  {author} {\bibfnamefont {P.}~\bibnamefont {Zoller}},\ }\bibfield  {title}
  {\bibinfo {title} {Inseparability criterion for continuous variable
  systems},\ }\href {https://doi.org/10.1103/PhysRevLett.84.2722} {\bibfield
  {journal} {\bibinfo  {journal} {Phys. Rev. Lett.}\ }\textbf {\bibinfo
  {volume} {84}},\ \bibinfo {pages} {2722} (\bibinfo {year}
  {2000})}\BibitemShut {NoStop}%
\bibitem [{\citenamefont {Mancini}\ \emph {et~al.}(2002)\citenamefont
  {Mancini}, \citenamefont {Giovannetti}, \citenamefont {Vitali},\ and\
  \citenamefont {Tombesi}}]{Mancini&TombesiPRL2002_DGCZ_product}%
  \BibitemOpen
  \bibfield  {author} {\bibinfo {author} {\bibfnamefont {S.}~\bibnamefont
  {Mancini}}, \bibinfo {author} {\bibfnamefont {V.}~\bibnamefont
  {Giovannetti}}, \bibinfo {author} {\bibfnamefont {D.}~\bibnamefont
  {Vitali}},\ and\ \bibinfo {author} {\bibfnamefont {P.}~\bibnamefont
  {Tombesi}},\ }\bibfield  {title} {\bibinfo {title} {Entangling macroscopic
  oscillators exploiting radiation pressure},\ }\href@noop {} {\bibfield
  {journal} {\bibinfo  {journal} {Physical Review Letters}\ }\textbf {\bibinfo
  {volume} {88}},\ \bibinfo {pages} {120401} (\bibinfo {year}
  {2002})}\BibitemShut {NoStop}%
\bibitem [{\citenamefont {Raymer}\ \emph {et~al.}(2003)\citenamefont {Raymer},
  \citenamefont {Funk}, \citenamefont {Sanders},\ and\ \citenamefont
  {De~Guise}}]{RaymerPRA2003}%
  \BibitemOpen
  \bibfield  {author} {\bibinfo {author} {\bibfnamefont {M.~G.}\ \bibnamefont
  {Raymer}}, \bibinfo {author} {\bibfnamefont {A.}~\bibnamefont {Funk}},
  \bibinfo {author} {\bibfnamefont {B.}~\bibnamefont {Sanders}},\ and\ \bibinfo
  {author} {\bibfnamefont {H.}~\bibnamefont {De~Guise}},\ }\bibfield  {title}
  {\bibinfo {title} {Separability criterion for separate quantum systems},\
  }\href@noop {} {\bibfield  {journal} {\bibinfo  {journal} {Physical Review
  A}\ }\textbf {\bibinfo {volume} {67}},\ \bibinfo {pages} {052104} (\bibinfo
  {year} {2003})}\BibitemShut {NoStop}%
\bibitem [{\citenamefont {Hillery}\ and\ \citenamefont
  {Zubairy}(2006{\natexlab{b}})}]{Hillery&ZubairyPRL2006}%
  \BibitemOpen
  \bibfield  {author} {\bibinfo {author} {\bibfnamefont {M.}~\bibnamefont
  {Hillery}}\ and\ \bibinfo {author} {\bibfnamefont {M.}~\bibnamefont
  {Zubairy}},\ }\bibfield  {title} {\bibinfo {title} {Entanglement conditions
  for two-mode states},\ }\href {https://doi.org/10.1103/PhysRevLett.96.050503}
  {\bibfield  {journal} {\bibinfo  {journal} {Phys. Rev. Lett.}\ }\textbf
  {\bibinfo {volume} {96}},\ \bibinfo {pages} {050503} (\bibinfo {year}
  {2006}{\natexlab{b}})}\BibitemShut {NoStop}%
\bibitem [{\citenamefont {Peres}(1996)}]{Peres:96}%
  \BibitemOpen
  \bibfield  {author} {\bibinfo {author} {\bibfnamefont {A.}~\bibnamefont
  {Peres}},\ }\bibfield  {title} {\bibinfo {title} {Separability criterion for
  density matrices},\ }\href {https://doi.org/10.1103/PhysRevLett.77.1413}
  {\bibfield  {journal} {\bibinfo  {journal} {Phys. Rev. Lett.}\ }\textbf
  {\bibinfo {volume} {77}},\ \bibinfo {pages} {1413} (\bibinfo {year}
  {1996})}\BibitemShut {NoStop}%
\bibitem [{\citenamefont {Horodecki}\ \emph {et~al.}(1996)\citenamefont
  {Horodecki}, \citenamefont {Horodecki},\ and\ \citenamefont
  {Horodecki}}]{Horodecki:96}%
  \BibitemOpen
  \bibfield  {author} {\bibinfo {author} {\bibfnamefont {M.}~\bibnamefont
  {Horodecki}}, \bibinfo {author} {\bibfnamefont {P.}~\bibnamefont
  {Horodecki}},\ and\ \bibinfo {author} {\bibfnamefont {R.}~\bibnamefont
  {Horodecki}},\ }\bibfield  {title} {\bibinfo {title} {Separability of mixed
  states: necessary and sufficient conditions},\ }\href@noop {} {\bibfield
  {journal} {\bibinfo  {journal} {Phys. Lett. A}\ }\textbf {\bibinfo {volume}
  {223}},\ \bibinfo {pages} {1} (\bibinfo {year} {1996})}\BibitemShut {NoStop}%
\bibitem [{\citenamefont {Agarwal}\ and\ \citenamefont
  {Biswas}(2005)}]{Agarwal&BiswasNJP2005}%
  \BibitemOpen
  \bibfield  {author} {\bibinfo {author} {\bibfnamefont {G.}~\bibnamefont
  {Agarwal}}\ and\ \bibinfo {author} {\bibfnamefont {A.}~\bibnamefont
  {Biswas}},\ }\bibfield  {title} {\bibinfo {title} {Inseparability
  inequalities for higher order moments for bipartite systems},\ }\href@noop {}
  {\bibfield  {journal} {\bibinfo  {journal} {New Journal of Physics}\ }\textbf
  {\bibinfo {volume} {7}},\ \bibinfo {pages} {211} (\bibinfo {year}
  {2005})}\BibitemShut {NoStop}%
\bibitem [{\citenamefont {Nha}\ and\ \citenamefont
  {Kim}(2006)}]{NhaPRA2006Fock_states}%
  \BibitemOpen
  \bibfield  {author} {\bibinfo {author} {\bibfnamefont {H.}~\bibnamefont
  {Nha}}\ and\ \bibinfo {author} {\bibfnamefont {J.}~\bibnamefont {Kim}},\
  }\bibfield  {title} {\bibinfo {title} {Entanglement criteria via the
  uncertainty relations in su (2) and su (1, 1) algebras: Detection of
  non-gaussian entangled states},\ }\href@noop {} {\bibfield  {journal}
  {\bibinfo  {journal} {Physical Review A}\ }\textbf {\bibinfo {volume} {74}},\
  \bibinfo {pages} {012317} (\bibinfo {year} {2006})}\BibitemShut {NoStop}%
\bibitem [{\citenamefont {Nha}\ and\ \citenamefont
  {Zubairy}(2008)}]{Nha&ZubairyPRL2008}%
  \BibitemOpen
  \bibfield  {author} {\bibinfo {author} {\bibfnamefont {H.}~\bibnamefont
  {Nha}}\ and\ \bibinfo {author} {\bibfnamefont {M.~S.}\ \bibnamefont
  {Zubairy}},\ }\bibfield  {title} {\bibinfo {title} {Uncertainty inequalities
  as entanglement criteria for negative partial-transpose states},\ }\href@noop
  {} {\bibfield  {journal} {\bibinfo  {journal} {Physical Review Letters}\
  }\textbf {\bibinfo {volume} {101}},\ \bibinfo {pages} {130402} (\bibinfo
  {year} {2008})}\BibitemShut {NoStop}%
\bibitem [{\citenamefont {Nha}(2007)}]{nha2007entanglement}%
  \BibitemOpen
  \bibfield  {author} {\bibinfo {author} {\bibfnamefont {H.}~\bibnamefont
  {Nha}},\ }\bibfield  {title} {\bibinfo {title} {Entanglement condition via su
  (2) and su (1, 1) algebra using schr{\"o}dinger-robertson uncertainty
  relation},\ }\href@noop {} {\bibfield  {journal} {\bibinfo  {journal}
  {Physical Review A}\ }\textbf {\bibinfo {volume} {76}},\ \bibinfo {pages}
  {014305} (\bibinfo {year} {2007})}\BibitemShut {NoStop}%
\bibitem [{\citenamefont {Shchukin}\ and\ \citenamefont
  {Vogel}(2005)}]{ShchukinVogelPRL2005}%
  \BibitemOpen
  \bibfield  {author} {\bibinfo {author} {\bibfnamefont {E.}~\bibnamefont
  {Shchukin}}\ and\ \bibinfo {author} {\bibfnamefont {W.}~\bibnamefont
  {Vogel}},\ }\bibfield  {title} {\bibinfo {title} {Inseparability criteria for
  continuous bipartite quantum states},\ }\href@noop {} {\bibfield  {journal}
  {\bibinfo  {journal} {Physical Review Letters}\ }\textbf {\bibinfo {volume}
  {95}},\ \bibinfo {pages} {230502} (\bibinfo {year} {2005})}\BibitemShut
  {NoStop}%
\bibitem [{\citenamefont {Ivan}\ \emph {et~al.}(2006)\citenamefont {Ivan},
  \citenamefont {Mukunda},\ and\ \citenamefont {Simon}}]{ivan2006generation}%
  \BibitemOpen
  \bibfield  {author} {\bibinfo {author} {\bibfnamefont {J.~S.}\ \bibnamefont
  {Ivan}}, \bibinfo {author} {\bibfnamefont {N.}~\bibnamefont {Mukunda}},\ and\
  \bibinfo {author} {\bibfnamefont {R.}~\bibnamefont {Simon}},\ }\bibfield
  {title} {\bibinfo {title} {Generation of npt entanglement from nonclassical
  photon statistics},\ }\href@noop {} {\bibfield  {journal} {\bibinfo
  {journal} {arXiv preprint quant-ph/0603255}\ } (\bibinfo {year}
  {2006})}\BibitemShut {NoStop}%
\bibitem [{\citenamefont {Ivan}\ \emph {et~al.}(2012)\citenamefont {Ivan},
  \citenamefont {Mukunda},\ and\ \citenamefont {Simon}}]{ivan2012generation}%
  \BibitemOpen
  \bibfield  {author} {\bibinfo {author} {\bibfnamefont {J.~S.}\ \bibnamefont
  {Ivan}}, \bibinfo {author} {\bibfnamefont {N.}~\bibnamefont {Mukunda}},\ and\
  \bibinfo {author} {\bibfnamefont {R.}~\bibnamefont {Simon}},\ }\bibfield
  {title} {\bibinfo {title} {Generation and distillation of non-gaussian
  entanglement from nonclassical photon statistics},\ }\href@noop {} {\bibfield
   {journal} {\bibinfo  {journal} {Quantum information processing}\ }\textbf
  {\bibinfo {volume} {11}},\ \bibinfo {pages} {873} (\bibinfo {year}
  {2012})}\BibitemShut {NoStop}%
\bibitem [{\citenamefont {Simon}(2000)}]{SimonPRL2000}%
  \BibitemOpen
  \bibfield  {author} {\bibinfo {author} {\bibfnamefont {R.}~\bibnamefont
  {Simon}},\ }\bibfield  {title} {\bibinfo {title} {Peres-horodecki
  separability criterion for continuous variable systems},\ }\href
  {https://doi.org/10.1103/PhysRevLett.84.2726} {\bibfield  {journal} {\bibinfo
   {journal} {Phys. Rev. Lett.}\ }\textbf {\bibinfo {volume} {84}},\ \bibinfo
  {pages} {2726} (\bibinfo {year} {2000})}\BibitemShut {NoStop}%
\bibitem [{\citenamefont {Scully}\ and\ \citenamefont
  {Zubairy}(1997)}]{ScullyZubairyBook}%
  \BibitemOpen
  \bibfield  {author} {\bibinfo {author} {\bibfnamefont {M.~O.}\ \bibnamefont
  {Scully}}\ and\ \bibinfo {author} {\bibfnamefont {M.~S.}\ \bibnamefont
  {Zubairy}},\ }\href@noop {} {\emph {\bibinfo {title} {Quantum Optics}}}\
  (\bibinfo  {publisher} {Cambridge University Press},\ \bibinfo {address} {New
  York},\ \bibinfo {year} {1997})\BibitemShut {NoStop}%
\bibitem [{\citenamefont {Braunstein}\ and\ \citenamefont {van
  Loock}(2005)}]{Braunstein:05}%
  \BibitemOpen
  \bibfield  {author} {\bibinfo {author} {\bibfnamefont {S.~L.}\ \bibnamefont
  {Braunstein}}\ and\ \bibinfo {author} {\bibfnamefont {P.}~\bibnamefont {van
  Loock}},\ }\bibfield  {title} {\bibinfo {title} {Quantum information with
  continuous variables},\ }\href {https://doi.org/10.1103/RevModPhys.77.513}
  {\bibfield  {journal} {\bibinfo  {journal} {Rev. Mod. Phys.}\ }\textbf
  {\bibinfo {volume} {77}},\ \bibinfo {pages} {513} (\bibinfo {year}
  {2005})}\BibitemShut {NoStop}%
\bibitem [{\citenamefont {Simon}\ \emph {et~al.}(1994)\citenamefont {Simon},
  \citenamefont {Mukunda},\ and\ \citenamefont {Dutta}}]{simon1994quantum}%
  \BibitemOpen
  \bibfield  {author} {\bibinfo {author} {\bibfnamefont {R.}~\bibnamefont
  {Simon}}, \bibinfo {author} {\bibfnamefont {N.}~\bibnamefont {Mukunda}},\
  and\ \bibinfo {author} {\bibfnamefont {B.}~\bibnamefont {Dutta}},\ }\bibfield
   {title} {\bibinfo {title} {Quantum-noise matrix for multimode systems: U (n)
  invariance, squeezing, and normal forms},\ }\href@noop {} {\bibfield
  {journal} {\bibinfo  {journal} {Physical Review A}\ }\textbf {\bibinfo
  {volume} {49}},\ \bibinfo {pages} {1567} (\bibinfo {year}
  {1994})}\BibitemShut {NoStop}%
\bibitem [{\citenamefont {Lee}(1991)}]{lee1991measure}%
  \BibitemOpen
  \bibfield  {author} {\bibinfo {author} {\bibfnamefont {C.~T.}\ \bibnamefont
  {Lee}},\ }\bibfield  {title} {\bibinfo {title} {Measure of the
  nonclassicality of nonclassical states},\ }\href@noop {} {\bibfield
  {journal} {\bibinfo  {journal} {Physical Review A}\ }\textbf {\bibinfo
  {volume} {44}},\ \bibinfo {pages} {R2775} (\bibinfo {year}
  {1991})}\BibitemShut {NoStop}%
\bibitem [{\citenamefont {Kiesel}\ and\ \citenamefont
  {Vogel}(2010)}]{vogelPRA2010Filters}%
  \BibitemOpen
  \bibfield  {author} {\bibinfo {author} {\bibfnamefont {T.}~\bibnamefont
  {Kiesel}}\ and\ \bibinfo {author} {\bibfnamefont {W.}~\bibnamefont {Vogel}},\
  }\bibfield  {title} {\bibinfo {title} {Nonclassicality filters and
  quasiprobabilities},\ }\href@noop {} {\bibfield  {journal} {\bibinfo
  {journal} {Physical Review A}\ }\textbf {\bibinfo {volume} {82}},\ \bibinfo
  {pages} {032107} (\bibinfo {year} {2010})}\BibitemShut {NoStop}%
\bibitem [{\citenamefont {Adesso}\ \emph {et~al.}(2004)\citenamefont {Adesso},
  \citenamefont {Serafini},\ and\ \citenamefont
  {Illuminati}}]{adesso2004extremal}%
  \BibitemOpen
  \bibfield  {author} {\bibinfo {author} {\bibfnamefont {G.}~\bibnamefont
  {Adesso}}, \bibinfo {author} {\bibfnamefont {A.}~\bibnamefont {Serafini}},\
  and\ \bibinfo {author} {\bibfnamefont {F.}~\bibnamefont {Illuminati}},\
  }\bibfield  {title} {\bibinfo {title} {Extremal entanglement and mixedness in
  continuous variable systems},\ }\href@noop {} {\bibfield  {journal} {\bibinfo
   {journal} {Phys. Rev. A}\ }\textbf {\bibinfo {volume} {70}},\ \bibinfo
  {pages} {022318} (\bibinfo {year} {2004})}\BibitemShut {NoStop}%
\bibitem [{\citenamefont {{G. Vidal and R. F. Werner}}(2002)}]{Vidal:02}%
  \BibitemOpen
  \bibfield  {author} {\bibinfo {author} {\bibnamefont {{G. Vidal and R. F.
  Werner}}},\ }\href@noop {} {\bibfield  {journal} {\bibinfo  {journal} {Phys.
  Rev. A}\ }\textbf {\bibinfo {volume} {65}},\ \bibinfo {pages} {032314}
  (\bibinfo {year} {2002})}\BibitemShut {NoStop}%
\bibitem [{\citenamefont {Plenio}(2005)}]{plenio2005logarithmic}%
  \BibitemOpen
  \bibfield  {author} {\bibinfo {author} {\bibfnamefont {M.~B.}\ \bibnamefont
  {Plenio}},\ }\bibfield  {title} {\bibinfo {title} {Logarithmic negativity: A
  full entanglement monotone that is not convex},\ }\href@noop {} {\bibfield
  {journal} {\bibinfo  {journal} {Phys. Rev. Lett.}\ }\textbf {\bibinfo
  {volume} {95}},\ \bibinfo {pages} {090503} (\bibinfo {year}
  {2005})}\BibitemShut {NoStop}%
\bibitem [{\citenamefont {Li}\ \emph {et~al.}(2006)\citenamefont {Li},
  \citenamefont {Li},\ and\ \citenamefont {Yang}}]{Li:06}%
  \BibitemOpen
  \bibfield  {author} {\bibinfo {author} {\bibfnamefont {H.-R.}\ \bibnamefont
  {Li}}, \bibinfo {author} {\bibfnamefont {F.-L.}\ \bibnamefont {Li}},\ and\
  \bibinfo {author} {\bibfnamefont {Y.}~\bibnamefont {Yang}},\ }\bibfield
  {title} {\bibinfo {title} {Entangling two single-mode gaussian states by use
  of a beam splitter},\ }\href@noop {} {\bibfield  {journal} {\bibinfo
  {journal} {Chinese Physics}\ }\textbf {\bibinfo {volume} {15}},\ \bibinfo
  {pages} {2947} (\bibinfo {year} {2006})}\BibitemShut {NoStop}%
\bibitem [{\citenamefont {Marian}\ and\ \citenamefont
  {Marian}(2018)}]{Marian_2018_JPhysA}%
  \BibitemOpen
  \bibfield  {author} {\bibinfo {author} {\bibfnamefont {P.}~\bibnamefont
  {Marian}}\ and\ \bibinfo {author} {\bibfnamefont {T.~A.}\ \bibnamefont
  {Marian}},\ }\bibfield  {title} {\bibinfo {title}
  {Einstein{\textendash}podolsky{\textendash}rosen-like separability indicators
  for two-mode gaussian states},\ }\href
  {https://doi.org/10.1088/1751-8121/aa9fae} {\bibfield  {journal} {\bibinfo
  {journal} {Journal of Physics A: Mathematical and Theoretical}\ }\textbf
  {\bibinfo {volume} {51}},\ \bibinfo {pages} {065301} (\bibinfo {year}
  {2018})}\BibitemShut {NoStop}%
\bibitem [{\citenamefont {Vaccaro}\ and\ \citenamefont
  {Pegg}(1990)}]{PeggJModOpt1990Intelligent}%
  \BibitemOpen
  \bibfield  {author} {\bibinfo {author} {\bibfnamefont {J.}~\bibnamefont
  {Vaccaro}}\ and\ \bibinfo {author} {\bibfnamefont {D.}~\bibnamefont {Pegg}},\
  }\bibfield  {title} {\bibinfo {title} {Physical number-phase intelligent and
  minimum-uncertainty states of light},\ }\href@noop {} {\bibfield  {journal}
  {\bibinfo  {journal} {Journal of Modern Optics}\ }\textbf {\bibinfo {volume}
  {37}},\ \bibinfo {pages} {17} (\bibinfo {year} {1990})}\BibitemShut {NoStop}%
\bibitem [{\citenamefont {Kitagawa}\ and\ \citenamefont
  {Yamamoto}(1986)}]{KitagawaPRA1986numberphase}%
  \BibitemOpen
  \bibfield  {author} {\bibinfo {author} {\bibfnamefont {M.}~\bibnamefont
  {Kitagawa}}\ and\ \bibinfo {author} {\bibfnamefont {Y.}~\bibnamefont
  {Yamamoto}},\ }\bibfield  {title} {\bibinfo {title} {Number-phase
  minimum-uncertainty state with reduced number uncertainty in a kerr nonlinear
  interferometer},\ }\href@noop {} {\bibfield  {journal} {\bibinfo  {journal}
  {Physical Review A}\ }\textbf {\bibinfo {volume} {34}},\ \bibinfo {pages}
  {3974} (\bibinfo {year} {1986})}\BibitemShut {NoStop}%
\end{thebibliography}%
  
\end{document}